\documentclass[12pt,onecolumn,draftclsnofoot]{IEEEtran}
\usepackage{ifpdf}
\usepackage{cite}
\ifCLASSINFOpdf
   \usepackage[pdftex]{graphicx}
\else
   \usepackage[dvips]{graphicx}
\fi
\usepackage{amsmath}
\usepackage{algorithmic}
\usepackage{array}
\ifCLASSOPTIONcompsoc
  \usepackage[caption=false,font=normalsize,labelfont=sf,textfont=sf]{subfig}
\else
  \usepackage[caption=false,font=footnotesize]{subfig}
\fi
\usepackage{fixltx2e}

\usepackage{bbm}
\usepackage{amssymb}
\usepackage{amscd}
\usepackage[amsmath,thref,thmmarks]{ntheorem}
\theoremstyle{plain}
\newtheorem{proposition}{Proposition}
\newtheorem{theorem}{Theorem}
\newtheorem{definition}{Definition}
\newtheorem{lemma}{Lemma}

\usepackage{hyperref}

\begin{document}
\title{Secure Polar Coding with Delayed Wiretapping Information}

\author{Yizhi~Zhao~and~Hongmei~Chi
        % <-this % stops a space
\thanks{Y. Zhao was with the College of Informatics, Huazhong Agricultural University, Wuhan,
Hubei, China. E-mail: zhaoyz@mail.hzau.edu.cn.}% <-this % stops a space
\thanks{H. Chi  was with the College of Science, Huazhong Agricultural University, Wuhan,
Hubei, China. E-mail: chihongmei@mail.hzau.edu.cn.}% <-this % stops a space
}

\maketitle

\begin{abstract}
In this paper, we investigate the secure coding issue for a wiretap channel model with fixed main channel and varying wiretap channel, by assuming that legitimate parties can obtain the wiretapping channel state information (CSI) after some time delay. For the symmetric degraded delay CSI case, we present an explicit weak security scheme by constructing secure polar codes on a one-time pad chaining structure, and prove its weak security, reliability and capability of approaching the secrecy capacity of perfect CSI case with delay CSI assumption. Further for the symmetric no-degraded delay CSI case, we present a modified multi-block chaining structure in which the original subset of frozen bit is designed for conveying functional random bits securely. Then we combine this modified multi-block chaining structure with the weak security scheme to construct an explicit strong security polar coding scheme, and prove its strong security, reliability and also the capability of approaching the secrecy capacity of perfect CSI case with delay CSI assumption. At last, we carry out stimulations to prove the performance of both secure schemes.
\end{abstract}


% \fi
\IEEEpeerreviewmaketitle

\section{Introduction}

\subsection{The Delay CSI Assumption}
\IEEEPARstart{P}{hysical} layer secure coding is an important and effective approach for secure and reliable communication over the \emph{wiretap channel (WTC)}\cite{Wyner1975}. In the last decade, after the invention of polar code\cite{Arikan2009}, secure polar coding schemes had successfully achieved the secrecy capacities of Wyner's wiretap channel\cite{Mahdavifar2011,Vard2013strong} and several extended wiretap channel models\cite{Hassani2014,Wei2015,Zheng2017,Chou2015,Chou2016,Wei2016,Gulcu2015}. All these studies are based on an idealized \emph{perfect channel state information (CSI)} assumption that channels of the WTC model are fixed and known by the legitimate parties during the entire communication process.

However, in practical communication there are always limitation and uncertainties of CSI due to realistic reasons such as estimation inaccuracy or eavesdropper's initiative\cite{Schaefer2015}. Since perfect CSI cannot hold for practical concerns, WTC models with uncertain CSI are proposed and studied. For instance, the \emph{compound wiretap channel}\cite{Liang2009} and the \emph{arbitrarily varying wiretap channel}\cite{Goldfeld2016} are the two typical uncertain CSI WTC models and their secrecy capacities are characterized in \cite{Bjelakovic2013,Bjelakovic2013av}. Comparing with the perfect CSI, secure coding with uncertain CSI is much more difficult.  In \cite{Si2015,Si2016}, a hierarchical polar coding has been proposed which can achieve the secrecy capacity without any instantaneous CSI. However, this technique can only be applied in the case of block fading channels that the CSI is varying with degraded relations and keep constant within each blocks. In \cite{Chou2018}, an explicit secure polar coding has been proposed which can achieve the lower bounds of the secrecy capacities for both compound wiretap channel and arbitrarily varying wiretap channel with fixed and publicly known main channels. But achieving the upper bounds of these secrecy capacities are still open problems.

Although the actual CSI cannot be known instantly in practical uncertain CSI cases, there are feasible approaches to obtain it \emph{with delay}. Existing studies have already begun to consider such delay cases. In \cite{Dai2017}, a delay CSI model has been studied that varying state is sent back to the legitimate transmitter by the legitimate receiver through a feedback channel after some time delay. In \cite{Tahmasbi2017}, a detectable assumption has been presented that legitimate parties can detect the physical effect in the environment caused by the varying of CSI and then learn the CSI from the detected information with high probability. In \cite{Wang2018}, deep learning algorithms are employed to learn the CSI of the time-varying massive MIMO channels from the feedback information. Accordingly, study the delay CSI could be a possible direction for the uncertain CSI problem.

Therefore, in this paper we setup a specific \emph{delay CSI assumption} for the WTC model as follow:
\begin{itemize}
\item Main channel is \emph{fixed} and \emph{publicly known}.
\item Wiretap channel can be either \emph{block varying} (similar as the compound wiretap channel model) or \emph{arbitrarily varying}.
\item Main channel and all the possible wiretap channel states are \emph{symmetric discrete memoryless channels without necessarily degraded relations}.
\item Legitimate parties can accurately obtain the wiretap channel state information only after $N$ times channel transmission, thus we can divide the entire transmission by $N$-length channel blocks.
\item We \emph{do not} assume that legitimate parties know the uncertain set of CSI or decoder know the CSI when decoding.
\end{itemize}

\subsection{Contributions of This Paper and Organization}

The objective of this paper is to investigate explicit information theoretical secure and reliable codes for the delay CSI WTC model. Since by our delay CSI assumption the main channel is fixed, achieving the reliability alone is not difficult. The real problem is to achieve reliability and security simultaneously without knowing the current wiretapping CSI over the entire encoding and decoding process, although the CSI can be obtained after.

First, we have constructed a secure polar coding scheme based on the \emph{one-time pad (OTP) chaining structure} of \cite{Tahmasbi2017}. To polarize the channel blocks, we use standard channel polarization on block varying model, and irregular channel polarization on arbitrarily varying model. Since this OTP chaining method has a similar sub-channel index partition structure as the secure polar coding of perfect CSI case, we can combine them together to form the secure polar coding scheme for the delay CSI cases. Theoretical analysis shows that this combined secure scheme achieves reliability and weak security in degraded delay CSI cases. However, it fails to achieve strong security in either degraded or non-degraded delay CSI cases.

Then based on the weak security scheme, we further construct a strong security polar coding scheme for the non-degraded delay CSI cases. Because the original \emph{multi-block chaining structure}\cite{Vard2013strong} cannot applied in the delay CSI cases, we have proposed a new modified multi-block chaining structure, in which part of the original frozen bit subset is designed for conveying functional random bits securely for the unreliable and insecure polarized sub-channels. Theocratical analysis indicates that the proposed strong security polar coding scheme achieves both reliability and strong security in non-degraded delay CSI cases. Also we have proven that the secrecy rate of our secure coding schemes can approach the secrecy capacity of the perfect CSI assumption in the delay CSI cases.

At last, we carry out stimulations to prove the performance of both secure schemes.

The outline of this paper is as follow. Section~\ref{sec_ps} presents the notations, communication models of delay CSI assumption and the main results of the paper. Section~\ref{sec_ws} presents the construction of a weak security polar coding scheme with discussions of the performance and remaining problems. Section~\ref{sec_ss} presents the construction of a strong security polar coding scheme with a modified multi-block chaining structure, and then analyzes its performance theoretically. Section~\ref{sec_simu} presents the simulation results of both weak and strong security schemes. Finally, Section~\ref{sec_con} concludes the paper.

\section{Problem Statements and Main Results}\label{sec_ps}

\subsection{Notations}

We define the integer interval $[\![a,b]\!]$ as the integer set between $\lfloor a\rfloor$ and $\lceil b\rceil$. For $n\in \mathbb{N}$, define $N\triangleq 2^n$. Denote $X$, $Y$, $Z$,... random variables (RVs) taking values in alphabets $\mathcal{X}$, $\mathcal{Y}$, $\mathcal{Z}$,... and the sample values of these RVs are denoted by $x$, $y$, $z$,... respectively. Then $p_{XY}$ denotes the joint probability of $X$ and $Y$, and $p_X$, $p_Y$ denotes the marginal probabilities. Also we denote vector $X^{a:b}\triangleq(X^a,X^{a+1},...,X^b)$ and when the context makes clear that we are dealing with vectors, we write $X^N$ in place of $X^{1:N}$ for simplification. And for any index set $\mathcal{A}\subseteqq [\![1,N]\!]$, we define $X^\mathcal{A}\triangleq \{X^i\}_{i\in \mathcal{A}}$. For the polar codes, we denote $\mathbf{G}_N$ the generator matrix , $\mathbf{R}$ the bit reverse matrix, $\mathbf{F}=
    \begin{bmatrix}\begin{smallmatrix}
        1 & 0 \\
        1 & 1
    \end{smallmatrix}\end{bmatrix}$
, $\otimes$ the Kronecker product, and have $\mathbf{G}_N=\mathbf{RF}^{\otimes n}$. $\mathbb{A}[\cdot]$ denotes the average.

\vbox{}

\subsection{Problem Statements}

In this paper we consider the scenario of our delay CSI assumption as follow: two legitimate users are communicating over a publicly known and fixed main channel while an eavesdropper is wiretapping through a block varying or arbitrarily varying wiretap channel; legitimate users do not know the current wiretapping state when encoding and decoding, but they can obtain the state after $N$ times channel transmission by certain approaches. For this delay CSI case, we first define the system model.

\begin{definition} The system model of delay CSI assumption is defined as $(\mathcal{X},\mathcal{Y},\mathcal{Z},\mathcal{S},p_{Y|X})$. $\mathcal{X}$ is the input alphabet of main channel $W$. $\mathcal{Y}$ is the output alphabet of main channel. $\mathcal{Z}$ is the output alphabet of the varying wiretap channel $V$. $\mathcal{S}$ is the set of potential wiretap channel states (uncertainty set). For each $s\in \mathcal{S}$, $s$ represents a potential transition probability of wiretap channel as $p_{Z|X}^{(s)}$ . $p_{Y|X}$ is the transition probability of main channel. For main channel and all potential wiretap channels, they are symmetric but with no necessarily degraded relations. Then for each $N$-length channel blocks, $\left(x^N,y^N,z^N\right)\in \mathcal{X}^N\times\mathcal{Y}^N \times \mathcal{Z}^N$, the main channel block have
\begin{equation}
p_{Y^N|X^N}\left(y^N|x^N\right)=\prod_{i=1}^{N}p_{Y|X}\left(y^i|x^i\right).
\label{eq_channeldefine}
\end{equation}
For the varying wiretap channel, we define two different sub-models:
\begin{itemize}
  \item block varying sub-model: denote $S_t$ the state of $t$-th wiretap channel block chosen by the eavesdropper from $\mathcal{S}$ with realization $s_t$. Then this state remains constant during the block communication. For the block transition probability, have
      \begin{equation}
      p^{(s_t)}_{Z^N|X^N}\left(z^N|x^N\right)=\prod_{i=1}^{N}p_{Z|X}^{(s_t)}\left(z^i|x^i\right);
      \end{equation}
  \item arbitrarily varying sub-model: denote $S_t^{1:N}$ the state of the $t$-th wiretap channel block chosen by the eavesdropper from $\mathcal{S}$ with realization $s_t^{1:N}$. Each $S_{t}^i$ represents the state for the $i$-th wiretap channel in the block. Then for the block transition probability, have
      \begin{equation}
      p^{(s_t^{1:N})}_{Z^N|X^N}\left(z^N|x^N\right)=\prod_{i=1}^{N}p_{Z|X}^{(s_{t}^i)}\left(z^i|x^i\right).
      \end{equation}
\end{itemize}
For legitimate parties, they can know the precise CSI only after each block communication. Besides, we use $\mathbf{s}$ to represent the CSI realization of both block varying and arbitrarily varying cases, that
\begin{equation}\label{eq_s}
\mathbf{s}=
\begin{cases}
  s, & \mbox{if block varying case}; \\
  s^{1:N}, & \mbox{if arbitrarily varying case}.
\end{cases}
\end{equation}
\end{definition}

\begin{figure}[!h]
\centering
\includegraphics[width=12cm]{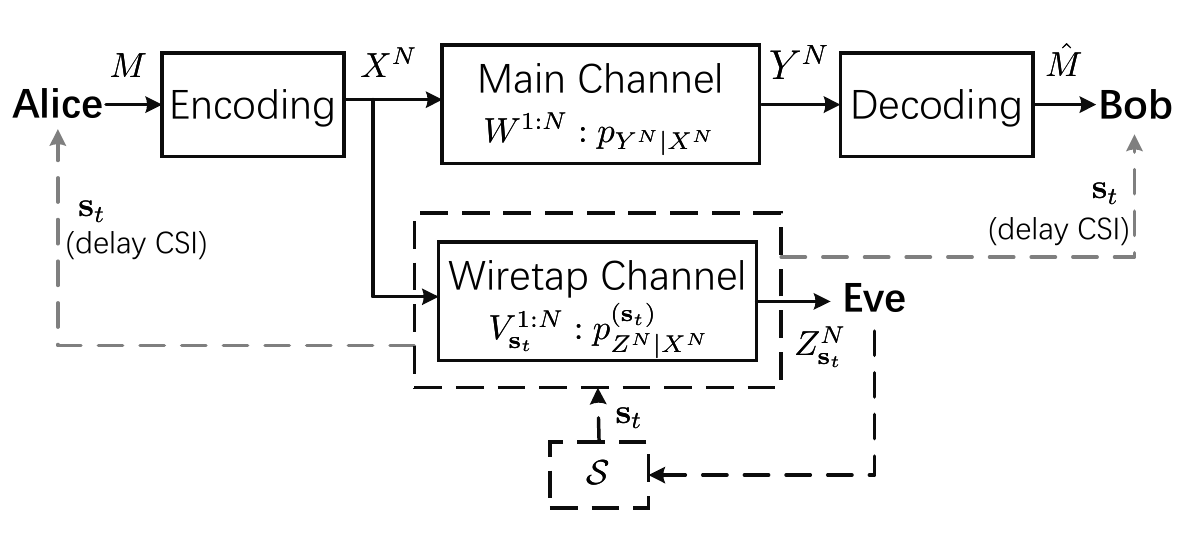}
\caption{The communication model of delay CSI assumption.}
\label{fig_vwtc}
\end{figure}

The communication process of the system model for $t$-th channel block is illustrated in Fig.~\ref{fig_vwtc}.
\begin{itemize}
\item Eavesdropper Eve chooses the CSI of the block as $\mathbf{s}_t$ and then obtains $Z^N_{\mathbf{s}_t}$ from her chosen wiretap channel with $p_{Z^N|X^N}^{(\mathbf{s}_t)}$ .
\item Legitimate transmitter Alice encodes the message $M$ into $X^N$ and transmits it to Bob over the main channel with $p_{Y^N|X^N}$, but she does not know the actual CSI $\mathbf{s}_t$  for the wiretap channel only until the end of $t$-th block communication.
\item Legitimate receiver Bob receives the main channel outputs $Y^N$ and decodes it into $\hat{M}$. He also does not know the actual CSI $\mathbf{s}_t$ for the wiretap channel only until the end of $t$-th block communication.
\end{itemize}

\begin{definition}(Performance)
Consider a $(2^{NR},N)$ code for the communication model of the delay CSI assumption, then the performance of this code can be measured by
\begin{itemize}
\item error probability:
\begin{equation}
\mathrm{P_e}=\Pr(M\neq \widehat{M});
\label{eq_errorprob}
\end{equation}
\item information leakage to Eve:
\begin{equation}
\mathrm{L}=I(Z^N_{\mathbf{s}_t};M).
\label{eq_leakage}
\end{equation}
\end{itemize}
\end{definition}

\begin{definition}(Criterions)
A rate $R$ is achievable if sequence of $(2^{NR},N)$ code exists under the criterions listed below:
\begin{itemize}
\item reliability criterion:
\begin{equation}
\lim\limits_{N\rightarrow\infty}\mathrm{P_e}=0;
\label{eq_reliability}
\end{equation}
\item weak security criterion:
\begin{equation}
\lim\limits_{N\rightarrow\infty}\frac{1}{N}\mathrm{L}=0;
\label{eq_weak}
\end{equation}
\item strong security criterion:
\begin{equation}
\lim\limits_{N\rightarrow\infty}\mathrm{L}=0.
\label{eq_strong}
\end{equation}
\end{itemize}
For multi state cases, above criterions take the maximum over all possible channel realizations.
\label{def_criterion}
\end{definition}

Then based on the delay CSI model defined above, our aim is to construct explicit secure polar codes to achieve a secure and reliable communication. Particularly we consider the information theoretical security and reliability that our secure polar codes have to achieve both reliability criterion and security criterion of Definition~\ref{def_criterion}.

\subsection{Main Results}

Considering a perfect CSI case of our varying WTC model that legitimate parties can directly know the precise CSI when the wiretapping state is varying. Then the system model equals to a basic WTC model. Thus according to the capacity result in \cite{Csiszar1978}, for any CSI realization $s\in\mathcal{S}$ in this perfect CSI case, the current secrecy capacity of current channels is
\begin{equation}
\mathrm{C_{s-perfect}}=\max_{V-X-Y,Z_s}\left[I\left(V;Y\right)-I\left(V;Z_s\right)\right].
\label{eq_perfect_c}
\end{equation}
In case of block varying, the state realization for the block remains constant as $s\in\mathcal{S}$, so the average secrecy capacity of the $N$ length block with perfect CSI is
\begin{equation}
\begin{split}
\mathrm{C_{s-Bperfect}}&=\frac{1}{N}\max_{V^N - X^N - Y^N,Z^N_{s}}\left[I(V^N;Y^N)-I(V^N;Z^N_{s})\right]\\
&=\max_{V-X-Y,Z_s}\left[I\left(V;Y\right)-I\left(V;Z_s\right)\right];
\end{split}
\label{eq_perfect_cb}
\end{equation}
In case of arbitrarily varying, the state realization of the block is $s^{1:N}\in\mathcal{S}^N$, so the average secrecy capacity of the $N$ length block with perfect CSI is
\begin{equation}\label{eq_perfect_ca}
\mathrm{C_{s-Aperfect}}=\frac{1}{N}\max_{V^N - X^N - Y^N,Z^N_{s^{1:N}}}\left[I(V^N;Y^N)-I(V^N;Z^N_{s^{1:N}})\right]
\end{equation}

For delay CSI cases, let $\mathrm{C_{s-Bdelay}}$ be the average secrecy capacity for block varying case and $\mathrm{C_{s-Adelay}}$ for arbitrarily case. Then we can observe that
\begin{equation}
\mathrm{C_{s-Bdelay}}\leq\mathrm{C_{s-Bperfect}},~\mathrm{C_{s-Adelay}}\leq\mathrm{C_{s-Aperfect}}.
\end{equation}

Note that in our  delay CSI WTC model, all channels are set symmetric. This is because in delay CSI case it is almost impossible to know the optimal distribution of a asymmetric channel for achieving the asymmetric secrecy capacity. In fact if we could know the optimal distribution, we can directly add the mature asymmetric channel polar coding technique of \cite{Arikan2010,Honda2013,Chou2015arg} in our proposed secure coding scheme, which barely changes the structure.

Then here is the \textbf{\emph{main results}} of the paper:
\begin{itemize}
  \item For the \emph{degraded} delay CSI WTC model, we propose a \emph{weak} security polar coding scheme which satisfies $\lim_{N\rightarrow\infty}\mathrm{P_e}=0$ and $\lim_{N\rightarrow\infty}\mathrm{L}/N=0$. And with delayed CSI, the secrecy rate $\mathrm{R_s}$ of the weak security scheme approaches the secrecy capacity of perfect CSI in \eqref{eq_perfect_cb} for block varying case and \eqref{eq_perfect_ca} for arbitrarily varying case.
  \item For the \emph{non-degraded} delay CSI WTC model, we propose a \emph{strong} security polar coding scheme which satisfies $\lim_{N\rightarrow\infty}\mathrm{P_e}=0$ and $\lim_{N\rightarrow\infty}\mathrm{L}=0$. And with delayed CSI, the secrecy rate $\mathrm{R_s}$ of the strong security scheme also approaches the secrecy capacity of perfect CSI in \eqref{eq_perfect_cb} for block varying case and \eqref{eq_perfect_ca} for arbitrarily varying case.
  \item We carry out stimulations to prove the performance of both secure schemes.
\end{itemize}

\section{Weak Security Polar Coding Scheme}\label{sec_ws}

In this section, we construct a secure polar coding scheme based on the one-time pad (OTP) chaining structure in \cite{Tahmasbi2017} for the delay CSI WTC model with both block varying and arbitrarily varying cases. Then we theoretically analyze its performance of the scheme and discuss the remaining problems.

\subsection{Polarized Subsets Partition}

First we present the channel polarization for both main channel and wiretap channel.

\begin{definition}(Bhattaharyya parameter)
Consider a pair of random variables $(X,Y)\sim p_{XY}$, where $X$ is a binary random variable and $Y$ is a finite-alphabet random variable. To measure the amount of randomness in $X$ given $Y$, the Bhattaharyya parameter is defined as
\begin{equation}
Z(X|Y)=2\sum_{y\in \mathcal{Y}}p_{Y}(y)\sqrt{p_{X|Y}(0|y)p_{X|Y}(1|y)}.
\label{eq_bhatdefine}
\end{equation}
\end{definition}

For the main channel, we have the publicly know and fixed channel block $W^{1:N}$ with same transition probability $p_{Y|X}$. Then according to the standard channel polarization theory in \cite{Arikan2009}, the main channel block can be polarized into almost full noise subset $\mathcal{H}_{X|Y}$ and almost noiseless subset $\mathcal{L}_{X|Y}$, that for $\beta\in\left(0,1/2\right)$, $\delta_{N}=2^{-N^\beta}$, have
\begin{equation}
\begin{split}
&\mathcal{H}_{X|Y}=\left\{i\in[\![1,N]\!]:Z\left(U^i|U^{1:i-1},Y^N\right)\geq1-\delta_N \right\}\\
&\mathcal{L}_{X|Y}=\left\{i\in[\![1,N]\!]:Z\left(U^i|U^{1:i-1},Y^N\right)\leq\delta_N \right\}.
\end{split}
\end{equation}

Then for the wiretap channel, we respectively consider the block varying case and arbitrarily varying case.

In the case of block varying, fory $s\in \mathcal{S}$, we have the wiretap channel block as $V_s^{1:N}$ with same transition probability $p_{Z|X}^{(s)}$. Then according to the standard channel polarization theory, the wiretap channel block $V_s^{1:N}$ can be polarized into almost full noise subset $\mathcal{H}_{X|Z}^{(s)}$ and almost noiseless subset $\mathcal{L}_{X|Z}^{(s)}$, that for $\beta\in\left(0,1/2\right)$, $\delta_{N}=2^{-N^\beta}$, have
\begin{equation}
\begin{split}
&\mathcal{H}_{X|Z}^{(s)}=\left\{i\in[\![1,N]\!]:Z\left(U^i|U^{1:i-1},Z^N_{s}\right)\geq1-\delta_N \right\}\\
&\mathcal{L}_{X|Z}^{(s)}=\left\{i\in[\![1,N]\!]:Z\left(U^i|U^{1:i-1},Z^N_{s}\right)\leq\delta_N \right\}.
\end{split}
\end{equation}

In the case of arbitrarily varying, we have the states $s^{1:N}\in \mathcal{S}^N$ for the wiretap channel block as $V_{s^{1:N}}^{1:N}$ with transition probability $p_{Z|X}^{(s^i)}$ for the $i$-th channel $V_{s^i}^i$. Note that in the arbitrarily varying case, the wiretap channel block is non-stationary, so that the basic channel polarization theory cannot be applied. The polarization of such non-stationary channels has been first proofed in \cite{Alsan2014} and further studied in \cite{Mahdavifar2018,Zhao2018}.

\begin{theorem}\label{theorem_polarization} (\cite{Zhao2018}) for any non-stationary B-DMC block $W^{1:N}$, by applying the irregular channel transformation $\mathbf{G}_N$, the generated channels ${W_N^{i}}$ can be polarized in the sense that, for any fixed $\delta\in(0,1)$, as $N\rightarrow\infty$, the fraction of indices $i\in[\![1,N]\!]$  for which $I(W_N^{i})\in (1-\delta,1]$ goes to $\mathbb{A}[I(W^{1:N })]$ and the fraction for which $I(W_N^{i})\in [0,\delta)$ goes to $1-\mathbb{A}[I(W^{1:N })]$. Also can be write as
\begin{equation}
I_\infty=
\begin{cases}
1 &~\text{w.p.} ~\mathbb{A}[I(W^{1:N })]\\
0 &~\text{w.p.}~1-\mathbb{A}[I(W^{1:N })]
\end{cases}
\end{equation}
where $\mathbb{A}[I(W^{1:N })]$ is the average of the initial $I(W^i)$ for all the $i\in[\![1,N]\!]$.
\end{theorem}

\begin{theorem}\label{theorem_rate}(\cite{Zhao2018}) For any non-stationary B-DMC blocks $W^{1:N}$ with $I(W^i)\geq0$, and any fixed $R<\mathbb{A}[I(W^{1:N})]$ and constant $\beta<1/2$, there exists index set $\mathcal{A}_N\subset[\![1,N]\!]$, $|\mathcal{A}_N|\leq NR$ that
\begin{equation}
\sum_{i\in\mathcal{A}_N}Z(W_N^{(i)})=o(2^{-N^\beta})
\end{equation}
and
\begin{equation}
\mathrm{P_e}(N,R)=o(2^{-N^\beta})
\end{equation}
\end{theorem}

Thus according to the irregular channel polarization theory, the non-stationary wiretap channel block $V_{s^{1:N}}^{1:N}$ of arbitrarily varying case can be polarized into almost full noise subset $\mathcal{H}_{X|Z}^{(s^N)}$ and almost noiseless subset $\mathcal{L}_{X|Z}^{(s^N)}$, that for $\beta\in\left(0,1/2\right)$, $\delta_{N}=2^{-N^\beta}$, have
\begin{equation}
\begin{split}
&\mathcal{H}_{X|Z}^{(s^{1:N})}=\left\{i\in[\![1,N]\!]:Z\left(U^i|U^{1:i-1},Z^N_{s^{1:N}}\right)\geq1-\delta_N \right\}\\
&\mathcal{L}_{X|Z}^{(s^{1:N})}=\left\{i\in[\![1,N]\!]:Z\left(U^i|U^{1:i-1},Z^N_{s^{1:N}}\right)\leq\delta_N \right\}.
\end{split}
\end{equation}

Note the transition probability $p_{Y|X}$ of the main channel is known and fixed over the entire multi-block communication, so that legitimate parties directly have the polarization result of main channel. But they cannot know the polarization result of current wiretap channel block until they obtain the block state $\mathbf{s}$ by the end of current block communication. For eavesdropper Eve, she directly knows all the polarization results of main channel block and wiretap channel block.

Next based on the above channel polarization results with block CSI $\mathbf{s}$, defined in \eqref{eq_s}, we can divide the channel block index $[\![1,N]\!]$ for both block varying case and arbitrarily varying case as follow:
\begin{equation}
\begin{split}
&\mathcal{I}^{(\mathbf{s})}=\mathcal{L}_{X|Y}\cap \mathcal{H}_{X|Z}^{(\mathbf{s})}\\
&\mathcal{F}^{(\mathbf{s})}=\left(\mathcal{L}_{X|Y}\right)^c\cap \mathcal{H}_{X|Z}^{(\mathbf{s})}\\
&\mathcal{R}^{(\mathbf{s})}=\mathcal{L}_{X|Y}\cap \left(\mathcal{H}_{X|Z}^{(\mathbf{s})}\right)^c\\
&\mathcal{B}^{(\mathbf{s})}=\left(\mathcal{L}_{X|Y}\right)^c\cap \left(\mathcal{H}_{X|Z}^{(\mathbf{s})}\right)^c.
\end{split}
\label{eq_division}
\end{equation}

Note that for both block varying and arbitrarily varying cases, subset $\mathcal{I}^{(\mathbf{s})}$ is secure and reliable, subset $\mathcal{F}^{(\mathbf{s})}$ is secure but unreliable, subset $\mathcal{R}^{(\mathbf{s})}$ is reliable but insecure, subset $\mathcal{B}^{(\mathbf{s})}$ is neither secure nor reliable. Subset reliable for Bob is fixed to $\mathcal{L}_{X|Y}$, subset unreliable for Bob is fixed to $\left(\mathcal{L}_{X|Y}\right)^c$, and have
\begin{equation}
\begin{split}
&\mathcal{I}^{(\mathbf{s})}\cup \mathcal{R}^{(\mathbf{s})}=\mathcal{L}_{X|Y}\text{~and~}\mathcal{F}^{(\mathbf{s})}\cup \mathcal{B}^{(\mathbf{s})}=\left(\mathcal{L}_{X|Y}\right)^c\\
&\lim_{N\rightarrow\infty}\frac{1}{N}|\mathcal{I}^{(\mathbf{s})}\cup \mathcal{R}^{(\mathbf{s})}|=I(U;Y)\\
&\lim_{N\rightarrow\infty}\frac{1}{N}|\mathcal{R}^{(\mathbf{s})}\cup \mathcal{B}^{(\mathbf{s})}|=\lim_{N\rightarrow\infty}\frac{1}{N}I(U^N;Z^N_\mathbf{s}).
\end{split}
\label{eq_rmk1}
\end{equation}
And for degraded wiretap channel cases, have
\begin{equation}
\begin{split}
\lim_{N\rightarrow\infty}\frac{1}{N}|\mathcal{I}^{(\mathbf{s})}|&=\lim_{N\rightarrow\infty}\frac{1}{N}|\mathcal{L}_{X|Y}\cap \mathcal{H}_{X|Z}^{(\mathbf{s})}|\\
&=I(U;Y)-\lim_{N\rightarrow\infty}\frac{1}{N}I(U^N;Z_\mathbf{s}^N)
\end{split}
\label{eq_degrade_rate}
\end{equation}
and
\begin{equation}
\lim_{N\rightarrow\infty}\frac{1}{N}|\mathcal{B}^{(\mathbf{s})}|=\lim_{N\rightarrow\infty}\frac{1}{N}|\left(\mathcal{L}_{X|Y}\right)^c\cap \left(\mathcal{H}_{X|Z}^{(\mathbf{s})}\right)^c|=0.
\label{eq_degraded_cs}
\end{equation}

\subsection{OTP Chain Based Secure Polar Coding Scheme }

With the polarized subsets partition in \eqref{eq_division}, now we construct the secure polar coding scheme for the delay CSI case. The basic idea of our construction is to embed the OPT chaining structure\cite{Tahmasbi2017} in the secure polar coding scheme. Note that in the rest of the paper, we use simplified expression $(\mathcal{I},\mathcal{R},\mathcal{B},\mathcal{F})$ to represent the partition $(\mathcal{I}^{(\mathbf{s})},\mathcal{R}^{(\mathbf{s})},\mathcal{B}^{(\mathbf{s})},\mathcal{F}^{(\mathbf{s})})$.

\begin{figure*}[!h]
\centering
\includegraphics[width=16cm]{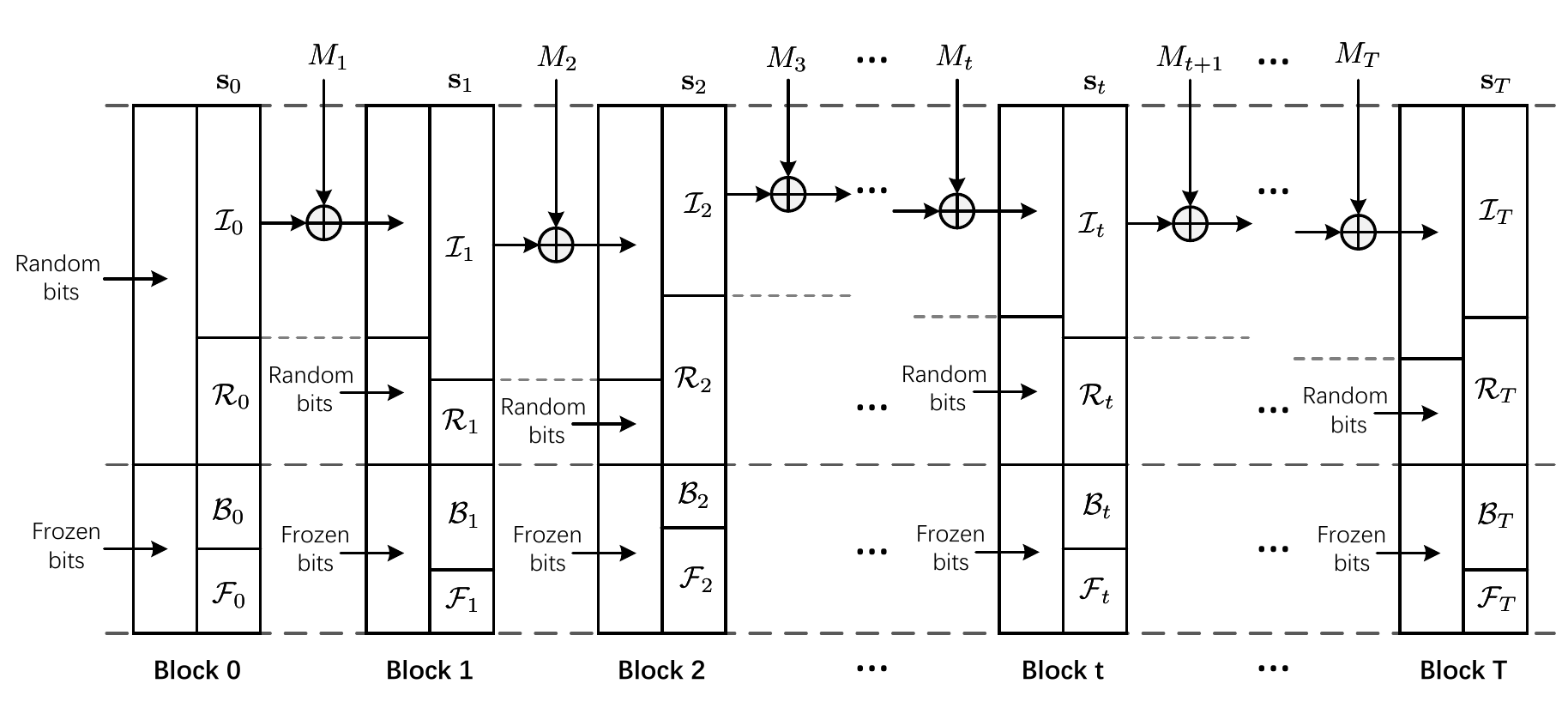}
\caption{The structure of OTP chain based secure polar coding scheme.}
\label{fig_pccs}
\end{figure*}

Fig.~\ref{fig_pccs} illustrates the entire structure of our proposed OTP chain based secure polar coding scheme which contains $T+1$ $N$-length block communication from block $0$ to block $T$. For block $t\in[\![0,T]\!]$ with the CSI realization $\mathbf{s}_t$, the actual partition of polarized subsets is denoted as $(\mathcal{I}_t,\mathcal{R}_t,\mathcal{B}_t,\mathcal{F}_t)$. For legitimate parties, since they do not have the CSI realization of the current wiretap channel at the point of encoding and decoding, they can only guarantee the reliability of the transmitted information by the fixed reliable polarized subset $\mathcal{L}_{X|Y}$ of the main channel. But every time when the current block communication is complete, they can accurately have the CSI realization $\mathbf{s}_t$ and then obtain the polarized subsets $(\mathcal{I}_t,\mathcal{R}_t,\mathcal{B}_t,\mathcal{F}_t)$. Therefore they can identify the bits that have just been reliably and securely transmitted, as the part in $\mathcal{I}_t$, and also the bits that have just been reliably but insecurely transmitted, as the part in $\mathcal{R}_t$.

According to the idea of the OTP chaining structure in \cite{Tahmasbi2017}, legitimate users can identify the secure and reliable part of the just completed communication block by the delay CSI and then use it as the key stream for the next block communication.

It seems that the polarized subset partition of secure polar codes can perfectly match the idea of the OTP chaining structure: bits in $\mathcal{I}_t$ are secure and reliable, so they can be used as the key stream to one-time-pad the message $M_{t+1}$ for $(t+1)$-th block communication by the legitimate transmitter, then the legitimate receiver can correctly decode the corresponding bits for decrypting the received message in $(t+1)$-th block communication.

Unfortunately, there are \textbf{\emph{serious flaws}} for this combined structure. In the non-degraded delay CSI cases, since the realization of CSI can only be obtained by the legitimate parties after each block communication, the unreliable and insecure subset $\mathcal{B}$ of the current block can not be identified from the unreliable subset $\left(\mathcal{L}_{X|Y}\right)^c$ at the time of encoding. \emph{Thus for reliability consideration, as a preliminary solution for the structure, subset $\mathcal{B}$ is assigned with the publicly known frozen bits together with the original frozen subset $\mathcal{F}$. However, this preliminary solution may compromise the security in non-degraded cases, which will be discussed in the next section.}

Now we present the construction of the OTP chain based secure polar coding scheme in Fig.~\ref{fig_pccs}. Note that polarized subsets $(\mathcal{H}_{X|Y},\mathcal{L}_{X|Y})$ remain constant during the entire $(T+1)$ times communication, and for any $t\in[\![1,T]\!]$, have $\mathcal{I}_{t-1}\subseteq \mathcal{L}_{X|Y}$, thus subset $\mathcal{I}_{t-1}$ is reliable for both $U_{t-1}^N$ and $U_t^N$, where $U_t^N$ represents the $U^N$ of $t$-th time block transmission

\vbox{}
\textbf{Block $0$:}
\begin{itemize}
\item Legitimate parties obtain the polarized subsets of the known and fixed main channel as $\mathcal{L}_{X|Y}$ and $\left(\mathcal{L}_{X|Y}\right)^c$;
\item Assigning the $u_0^N$ for polar coding:
\begin{itemize}
\item uniformly distributed random bits are assigned to the reliable subset $\mathcal{L}_{X|Y}$, also as $\mathcal{I}_0\cup\mathcal{R}_0$;
\item publicly known frozen bits are assigned to the unreliable subset $\left(\mathcal{L}_{X|Y}\right)^c$, also as $\mathcal{B}_0\cup\mathcal{F}_0$;
\end{itemize}
\item Alice encodes $u^N_0$ into the channel input $x^N_0$ by polar encoding $x^N_0=u^N_0\mathbf{G}_N$, and then transmits $x^N_0$ to Bob over the main channel block;
\item Bob receives $y^N_0$ from the main channel block and then decodes it into the estimated $\hat{u}^N_0$ by using the succussive cancelation (SC) decoding \cite{Arikan2009}:
\begin{equation}
\hat{u}^i_0=\begin{cases}
  \arg \max \limits_{u\in\left\{0,1\right\}}p_{U^i|U^{1:i-1}Y^N}\left(u|\hat{u}_0^{1:i-1}y^N_0\right)\text{, if } i\in\mathcal{L}_{X|Y} \\
  \text{publicly known frozen bit, if }i\in\left(\mathcal{L}_{X|Y}\right)^c
\end{cases}
\end{equation}
\item After the block communication, both Alice and Bob obtain the CSI of block $0$ as $\mathbf{s}_0$ and subset $\mathcal{I}_0$. Then Alice extracts $u^{\mathcal{I}_0}_0$ as the key stream for next block's encryption and Bob extracts $\hat{u}^{\mathcal{I}_0}_0$ as the key stream for next block's decryption.
\end{itemize}

\vbox{}
\textbf{Block $t$, $t\in[\![1,T]\!]$:}
\begin{itemize}
\item Legitimate parties obtain the divided subsets of last block as $(\mathcal{I}_{t-1},\mathcal{R}_{t-1},\mathcal{B}_{t-1},\mathcal{F}_{t-1})$;
\item Assume a binary message $M_t$ that satisfies $|M_t|=|\mathcal{I}_{t-1}|$. Then encrypt $|M_t|$ into ciphertext $E_t$ by $E_t=M_t\oplus u^{\mathcal{I}_{t-1}}_{t-1}$, where $\oplus$ is the XOR operation;
\item Assigning the $u_t^N$ for polar coding:
\begin{itemize}
\item ciphertext $E_t$ is assigned to subset $\mathcal{I}_{t-1}$;
\item uniformly distributed random bits are assigned to subset $\mathcal{R}_{t-1}$;
\item publicly known frozen bits are assigned to subset $\left(\mathcal{L}_{X|Y}\right)^c$;
\end{itemize}
\item Alice encodes $u^N_t$ into the optimally distributed channel input $x^N_t$ by polar encoding $x^N_t=u^N_t\mathbf{G}_N$, then transmits $x^N_t$ to Bob over the main channel block;
\item Bob receives $y^N_t$ from the main channel block and decodes it into the estimated $\hat{u}^N_t$ by using the SC decoding:
\begin{equation}
\hat{u}_t^i=
\begin{cases}
  \arg \max \limits_{u\in\left\{0,1\right\}}p_{U^i|U^{1:i-1}Y^N}\left(u|\hat{u}^{1:i-1}_ty^N_t\right)\text{, if } i\in\mathcal{L}_{X|Y} \\
  \text{publicly known frozen bit, if }i\in\left(\mathcal{L}_{X|Y}\right)^c
\end{cases}
\end{equation}
\item Bob extracts $\hat{u}^{\mathcal{I}_{t-1}}_t$ as the ciphertext and decrypts it by $\widehat{M}_t=\hat{u}^{\mathcal{I}_{t-1}}_t\oplus \hat{u}^{\mathcal{I}_{t-1}}_{t-1}$;
\item After the block communication, both Alice and Bob obtain the CSI of block $t$ as $\mathbf{s}_t$ and subset $\mathcal{I}_t$. Then Alice extracts $u^{\mathcal{I}_t}_t$ as the key stream for next block's encryption and Bob extracts $\hat{u}^{\mathcal{I}_t}_t$ as the key stream for next block's decryption.
\end{itemize}

\subsection{Performance Discussion}

Now we analyze the performance of the proposed OTP chain based secure polar coding scheme and discuss its existing problems.

In the scheme, ciphertext is carried by $U_t^{\mathcal{I}_{t-1}}$, and key stream is carried by $U_t^{\mathcal{I}_t}$, thus the reliability of the secure polar coding scheme is measured by the error probability of decoding the ciphertext and key stream from time $0$ to time $t$.

\begin{lemma}(\cite{Arikan2009})\label{lem_errorprob}
Considering an arbitrary subset $\mathcal{A}$ of block index $[\![1,N]\!]$ for DMC $W$, in case of $\mathcal{A}$ used as the information set and $\mathcal{A}^c$ used as frozen set for polar coding with
\begin{equation}
\mathcal{A}\subseteq \left\{i\in[\![1,N]\!]:Z\left(U^i|U^{1:i-1},Y^N\right)\leq\delta_N \right\},
\end{equation}
then for the successive cancellation decoding, $\beta\in\left(0,1/2\right)$, $\delta_{N}=2^{-N^\beta}$, the block error probability is bounded by
\begin{equation}
\mathrm{P_e}(\mathcal{A})\leq\sum_{i\in\mathcal{A}}Z(U^i|U^{1:i-1},Y^N)=O(2^{-N^\beta}).
\end{equation}
\end{lemma}

\begin{proposition}\label{prop_weak_reliability}
The proposed OTP chain based secure polar coding scheme, with setting $\mathcal{B}$ as frozen bit set, achieves reliability over the delay CSI WTC model.
\begin{IEEEproof}
In the entire $T+1$ times block communication, there are $T$ times ciphertext transmissions from time $1$ to time $T$, and $T$ times key stream transmissions from time $0$ to time $T-1$. Let $\mathrm{P_e}(T+1)$ be the decoding error probability of Bob for both ciphertext and key stream, have
\begin{equation}
\begin{split}
\mathrm{P_e}(T+1)=&\sum_{t=1}^T \sum_{i\in\mathcal{I}_{t-1}}Z(U^i|U^{1:{i-1}},Y^N)+\sum_{t=0}^{T-1} \sum_{i\in\mathcal{I}_t}Z(U^i|U^{1:{i-1}},Y^N)\\
\overset{(a)}{\leq}&TO(2^{-N^\beta})+TO(2^{-N^\beta})\\
=&2TO(2^{-N^\beta}),
\end{split}
\end{equation}
where $(a)$ is due to Lemma~\ref{lem_errorprob} and $(\mathcal{I}_t,\mathcal{I}_{t-1})\subseteq\mathcal{L}_{X|Y}$. Therefore the reliability can be achieved with a fixed $T$.
\end{IEEEproof}
\end{proposition}

Next we discuss the security of the polar coding based encrypted chaining structure under the reliability criterion.

\begin{lemma}\label{lem_fino} Considering a single block transmission with polar subset division in \eqref{eq_division} that $(U^\mathcal{I},U^{\mathcal{F}},U^{\mathcal{B}},U^{\mathcal{R}})- X^N- Z^N_\mathbf{s}$, in case that Eve have received $Z^N$ and knows $U^\mathcal{I}$, $U^{\mathcal{F}}$ and $U^{\mathcal{B}}$, then for $\beta\in\left(0,1/2\right)$, $\delta_{N}=2^{-N^\beta}$, have
\begin{equation}
H(U^{\mathcal{R}}|Z^N_\mathbf{s},U^\mathcal{I})\leq H(\delta_{N})+|\mathcal{R}|\delta_{N}
\end{equation}
\begin{IEEEproof} Define $\widehat{U}^{\mathcal{R}}=\mathbb{F}_{\mathrm{sc}}(Z^N_\mathbf{s},U^\mathcal{I})$ the SC decoding for Eve. Since $\mathcal{R}\subseteq \mathcal{L}_{X|Z}$, from Lemma~\ref{lem_errorprob}, have
\begin{equation}
\mathrm{P_e}(\text{Eve})=\Pr(U^{\mathcal{R}}\neq\widehat{U}^{\mathcal{R}})\leq O(2^{-N^\beta})
\end{equation}
Thus by applying the Fano's inequality, have
\begin{equation}
\begin{split}
H(U^{\mathcal{R}}|Z^N_\mathbf{s},U^\mathcal{I})&\leq H(\mathrm{P_e}(\text{Eve}))+|\mathcal{R}|\mathrm{P_e}(\text{Eve})\\
&=H(\delta_{N})+|\mathcal{R}|\delta_{N}
\end{split}
\end{equation}
\end{IEEEproof}
\end{lemma}

Note that in the structure, message are encrypted by one-time pad, thus the security can be measured by the information leakage of the key streams which are carried by $U^{\mathcal{I}_t}_t$. Let $\mathrm{L}_t$ be the information leakage of block $t$, then have $\mathrm{L}_t=I(U^{\mathcal{I}_t}_t;Z^N_t)$. Since for the entire $T+1$ blocks the transmission of key streams are independent between each blocks, the overall information leakage of key streams is $\sum_{t=0}^{T-1}\mathrm{L}_t$.

\begin{proposition} \label{prop_weak_security}The proposed OTP chain based secure polar coding scheme, with setting $\mathcal{B}$ as frozen bit set, achieves weak security over the degraded delay CSI WTC model, but fails to achieve strong security over either degraded or non-degraded delay CSI WTC models.
\begin{IEEEproof}
To simplify the expression, we omit most of the subscript $t$ in the following discussion. Note that in the structure, in order to maintain the reliability, subsets $U^{\mathcal{F}}$ and $U^{\mathcal{B}}$ are set for the publicly known frozen bits together. Therefore Eve can have the $U^{\mathcal{F}\cup\mathcal{B}}$ when she decodes the wiretapped message. Thus for the single block information leakage $\mathrm{L}_t$, have
\begin{equation}
\begin{split}
\mathrm{L}_t=&I(U^{\mathcal{I}};Z^N_\mathbf{s})\\
\overset{(a)}{=}&I(U^\mathcal{I},U^{\mathcal{F}\cup\mathcal{B}};Z^N_\mathbf{s})\\
=&I(U^{\mathcal{I}\cup\mathcal{F}\cup\mathcal{R}\cup\mathcal{B}};Z^N_\mathbf{s})-I(U^{\mathcal{R}};Z^N_\mathbf{s}|U^{\mathcal{I}\cup\mathcal{F}\cup\mathcal{B}})\\
=&I(U^N;Z^N_\mathbf{s})-I(U^{\mathcal{R}};Z^N_\mathbf{s}|U^\mathcal{I})\\
=&I(U^N;Z^N_\mathbf{s})-H(U^{\mathcal{R}})+H(U^{\mathcal{R}}|Z^N_\mathbf{s},U^\mathcal{I})\\
\overset{(b)}{\leq}&I(U^N;Z^N_\mathbf{s})-|\mathcal{R}|+H(\delta_{N})+|\mathcal{R}|\delta_{N},
\end{split}
\end{equation}
where $(a)$ is because $U^{\mathcal{F}\cup\mathcal{B}}$ is the publicly known frozen bits, $(b)$ is due to Lemma~\ref{lem_fino} and $U^{\mathcal{R}}$ are uniformly distributed random bits. From \eqref{eq_rmk1}, have
\begin{equation}
\begin{split}
\lim_{N\rightarrow\infty}\left[I(U^N;Z^N_\mathbf{s})-|\mathcal{R}|\right]&=\lim_{N\rightarrow\infty}\left[I(U^N;Z^N_\mathbf{s})-|\mathcal{R}\cup\mathcal{B}|+|\mathcal{B}|\right]\\
&=\lim_{N\rightarrow\infty}\left[I(U^N;Z^N_\mathbf{s})-I(U^N;Z^N_\mathbf{s})+|\mathcal{B}|\right]\\
&=\lim_{N\rightarrow\infty}|\mathcal{B}|.
\end{split}
\end{equation}
Thus we have
\begin{equation}
\lim_{N\rightarrow\infty}\mathrm{L}_t\leq\lim_{N\rightarrow\infty}|\mathcal{B}|\text{~and~}\lim_{N\rightarrow\infty}\frac{\mathrm{L}_t}{N}\leq\lim_{N\rightarrow\infty}\frac{|\mathcal{B}|}{N}.
\end{equation}

In the case of degraded delay CSI WTC model, we can have $\lim_{N\rightarrow\infty}\mathrm{L}_t/N=0$ by \eqref{eq_degraded_cs}, which implies that only the weak security can be achieved by the proposed scheme.

However in the case of non-degraded delay CSI WTC model, neither $|\mathcal{B}|$ nor $|\mathcal{B}|/N$ is vanishing when $N\rightarrow\infty$, thus the security criterions cannot be achieved for either the single block or the entire $T+1$ blocks.
\end{IEEEproof}
\end{proposition}

Since the proposed OTP chain based secure polar coding scheme achieves reliability and weak security over the degraded delay CSI WTC model, we analyze the corresponding achievable secrecy rate.

\begin{proposition} \label{prop_weak_rate}Over the degraded delay CSI WTC model, with a large enough $T$, the achievable secrecy rate of the proposed OTP chain based secure polar coding scheme can approach the secrecy capacity of perfect CSI case.
\begin{IEEEproof}
Let $\mathrm{R_s}(T+1)$ be the secrecy rate of entire $T+1$ block communication. Since in $t$-th block communication encrypted messages are transmitted in the subset $\mathcal{I}_{t-1}$, we have
\begin{equation}
\begin{split}
\lim_{N\rightarrow\infty}\mathrm{R_s}(T+1)&=\lim_{N\rightarrow\infty}\frac{1}{N(T+1)}\sum_{t=1}^{T}|\mathcal{I}_{t-1}|\\
&=\frac{1}{T+1}\sum_{t=0}^{T-1}\lim_{N\rightarrow\infty}\frac{|\mathcal{I}_t|}{N}\\
&\overset{(a)}=\frac{1}{T+1}\sum_{t=0}^{T-1}\left[ I(U;Y)-\lim_{N\rightarrow\infty}\frac{1}{N}I(U^N;Z_\mathbf{s}^N) \right],
\end{split}
\end{equation}
where $(a)$ is due to \eqref{eq_degrade_rate} of degraded wiretap channel cases. Then by comparing with \eqref{eq_perfect_cb} and \eqref{eq_perfect_ca}, we can observe that $I(U;Y)-\lim_{N\rightarrow\infty}\frac{1}{N}I(U^N;Z_\mathbf{s}^N)$ can reach the perfect CSI average secrecy capacity of a block with state $\mathbf{s}$. Thus the secrecy rate of $T+1$ blocks can approach the average secrecy capacity of perfect CSI case by choosing a large enough $T$.
\end{IEEEproof}
\end{proposition}

In the next section, we will discuss the remaining problem of subset $\mathcal{B}$ and explore a new strong security solution for the non-degraded delay CSI WTC model.

\section{Strong Security Polar Coding Scheme}\label{sec_ss}

As previously discussed, the proposed OTP chain based secure polar coding scheme fail to achieve strong security over the delay CSI WTC model because of the neither secure nor reliable subset $\mathcal{B}_t$. Thus in this section, we will present a new solution for this remaining problem and construct a modified secure polar coding scheme which can achieve strong security and reliability simultaneously.

\subsection{Further Discussions on Strong Security}

In our preliminary solution for subset $\mathcal{B}_t$ of the delay CSI WTC, $U^{\mathcal{B}_t}_t$ is assigned with publicly known frozen bits for achieving the reliability, which however has been proven for compromising the security.

For the non-degraded WTC with perfect CSI, this conflict between reliability and security has already been solved by the technique of polar code based multi-block chaining structure proposed in \cite{Vard2013strong}. \emph{The basic idea of this strong security solution is to convey the bits of $U^{\mathcal{B}}$ to legitimate receiver Bob separately while keeping it safe from the eavesdropper Eve.} Therefore, in the original multi-block chaining structure, for any channel block $t$, a reliable and secure subset $\mathcal{E}_t$ that satisfies $|\mathcal{E}_t|=|\mathcal{B}_{t+1}|$ is separated from the subset $\mathcal{I}_t$. Then $\mathcal{E}_t$ is set for carrying uniformly distributed random bits which will be used for assigning the subset $\mathcal{B}_{t+1}$ in block $t+1$. Thus when decoding, Bob can directly decode the bits in subset $\mathcal{B}_{t+1}$ by the decoded random bits of $\mathcal{E}_t$ from block $t$.

However, in the delay CSI WTC model, the original multi-block chaining structure cannot be applied. According to the delay CSI assumption, the subset $\mathcal{B}_{t+1}$ cannot be identified by the legitimate parties only until $(t+1)$-th block communication is complete. Thus without knowing the subset $\mathcal{B}_{t+1}$, bits in subset $\mathcal{B}_t$ can not be assigned independently from $\mathcal{F}_t$, and the corresponding subset $\mathcal{E}_t$ in $t$-th block communication cannot be constructed as well. Therefore, to achieve both strong security and reliability over the delay CSI WTC, we have solve this \emph{unidentifiable problem} of $\mathcal{B}_t$.

In fact, there is an easy way around this problem to achieve both reliability and security, but \emph{it will also cause unacceptable secrecy rate sacrifice.} According to the delay CSI assumption, although subset $\mathcal{B}_{t}$ is unidentifiable, subset $\left(\mathcal{L}_{X|Y}\right)^c$ is known and fixed over the blocks. Thus we can directly apply the multi-block chaining structure on the subset $\left(\mathcal{L}_{X|Y}\right)^c$ instead of the unknown subset $\mathcal{B}_{t}$. For example, considering the $\left(\mathcal{L}_{X|Y}\right)^c$ based multi-block chaining structure, for block $t$, construct a subset $\mathcal{E}_t$ from $\mathcal{I}_t$ that satisfies $|\mathcal{E}_t|=|\left(\mathcal{L}_{X|Y}\right)^c|$. Then for the achievable secrecy rate of block $t$, have
\begin{equation}
\begin{split}
\lim_{N\rightarrow\infty}\mathrm{R_s}&=\lim_{N\rightarrow\infty}\frac{1}{N}|\mathcal{I}\setminus \mathcal{E}|=\lim_{N\rightarrow\infty}\frac{1}{N}(|\mathcal{I}\cup \mathcal{R}|-|\mathcal{B}\cup \mathcal{R}|-|\mathcal{F}|)\\
&=I(U;Y)-I(U;Z)-\mathrm{R}_\mathcal{F}\\
&=\mathrm{C_s}-\mathrm{R}_\mathcal{F},
\end{split}
\label{eq_ratesacri}
\end{equation}
where $\mathrm{C_s}$ is the secrecy capacity of a single block, $\mathrm{R}_\mathcal{F}$ is the rate of subset $\mathcal{F}$. Thus as shown in \eqref{eq_ratesacri}, large part of the secrecy capacity is sacrificed if directly apply the multi-block chaining structure on the subset $\left(\mathcal{L}_{X|Y}\right)^c$.

\subsection{Modified Multi-block Chaining Structure}

In order to achieve the strong security and reliability, we have to find a method to convey random bits for the known and fixed $\left(\mathcal{L}_{X|Y}\right)^c$ in the delay CSI WTC model without unacceptable rate sacrifice. Thus in this subsection, we present a new solution for this problem named as \emph{modified multi-block chaining structure.}

Note that in our preliminary weak security solution, subset $\left(\mathcal{L}_{X|Y}\right)^c$ is set as publicly known frozen bits for maintaining reliability. But what if we use it to transmit random bits instead? Since $\mathcal{F}\subseteq \mathcal{H}_{X|Z}$, random bits in $\mathcal{F}$ is secure from eavesdropper Eve. And with a delay CSI, by the end of every block communication, legitimate parties can know the actual subset $\mathcal{F}$ and $\mathcal{B}$ for identifying the secure part of the random bits in $\mathcal{H}_{X|Z}$ in the just completed block communication. Based on this point, we can use the subset $\mathcal{F}$ to construct a modified multi-block chaining structure.

\begin{figure*}[!h]
\centering
\includegraphics[width=16cm]{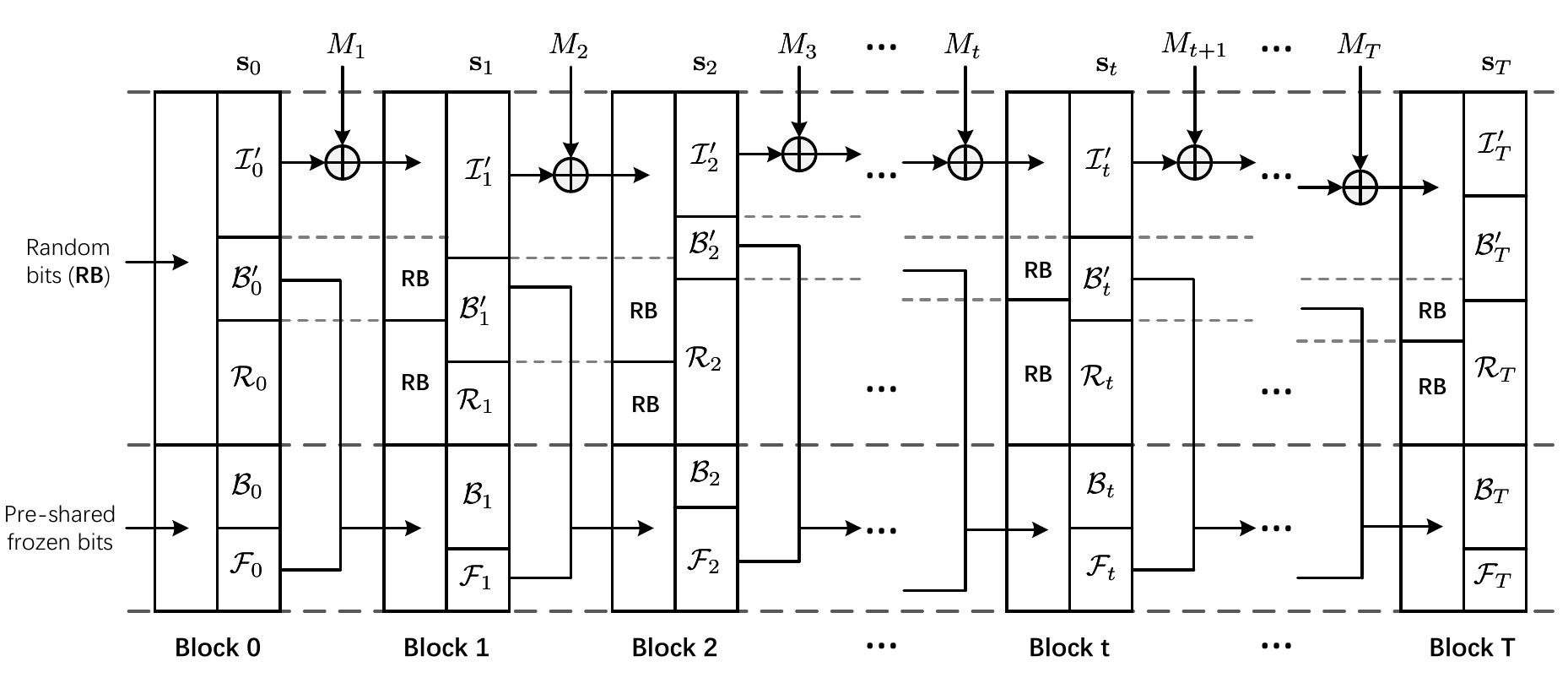}
\caption{The modified multi-block chaining structure for delay CSI assumption.}
\label{fig_fcs}
\end{figure*}

The modified multi-block chaining structure is illustrated in Fig.~\ref{fig_fcs}. Every time when legitimate parties obtain the CSI realization of the just completed block communication, they can know the actual divided subsets $(\mathcal{I},\mathcal{R},\mathcal{B},\mathcal{F})$ by \eqref{eq_division}. Assuming that over $\mathcal{S}$ have $|\mathcal{B}^{(\mathbf{s})}|<|\mathcal{I}^{(\mathbf{s})}|$ (if $|\mathcal{B}^{(\mathbf{s})}|\geq|\mathcal{I}^{(\mathbf{s})}|$, the secrecy capacity of state $\mathbf{s}$ is $0$), then we can divide the subset $\mathcal{I}$ into two parts, $\mathcal{B}'$ and $\mathcal{I}'$, which satisfies
\begin{equation}
\mathcal{B}'\subset \mathcal{I},~|\mathcal{B}'|=\mathcal{B}~\text{and}~\mathcal{I}'=\mathcal{I}\setminus \mathcal{B}'.
\end{equation}

At the beginning of time $0$, set a secure pre-shared frozen bits between Alice and Bob for assigning the $\left(\mathcal{L}_{X|Y}\right)^c$. Then random bits are assigned to $\mathcal{L}_{X|Y}$. When the block communication of time $0$ is completed, legitimate parties can obtain the CSI realization $s_0$. Accordingly they can identify the part of the pre-shared frozen bits that remains secure in the just completed transmission as the $U^{\mathcal{F}_0}_0$. Also they can identify the securely and reliably transmitted bit of $\mathcal{L}_{X|Y}$, as the $U^{\mathcal{B}'_0}_0$ and $U^{\mathcal{I}'_0}_0$ for Alice,  $\widehat{U}^{\mathcal{B}'_0}_0$ and $\widehat{U}^{\mathcal{I}'_0}_0$ for Bob.

Then for the block communication of time $1$, since $|\mathcal{F}_0|+|\mathcal{B}'_0|=|\left(\mathcal{L}_{X|Y}\right)^c|$, Alice can use the bits of $U^{\mathcal{F}_0}_0$ and $U^{\mathcal{B}'_0}_0$ together to assign the $U_1^{\left(\mathcal{L}_{X|Y}\right)^c}$. Since Bob already have the $U^{\mathcal{F}_0}_0$ and $\widehat{U}^{\mathcal{B}'_0}_0$, he can directly use these bits to decode $\widehat{U}_1^{\left(\mathcal{L}_{X|Y}\right)^c}$.

After the block communication of time $1$, legitimate parties can obtain the CSI realization $s_1$. Then Alice can identify the bits in $\left(\mathcal{L}_{X|Y}\right)^c$ that remains secure as $U^{\mathcal{F}_1}_1$, and for Bob as $\widehat{U}^{\mathcal{F}_1}_1$. Also they can identify the securely and reliably transmitted bit in $\mathcal{L}_{X|Y}$ as the $U^{\mathcal{B}'_1}_1$ and $U^{\mathcal{I}'_1}_1$ for Alice,  $\widehat{U}^{\mathcal{B}'_1}_1$ and $\widehat{U}^{\mathcal{I}'_1}_1$ for Bob. Then $(U^{\mathcal{F}_1}_1,U^{\mathcal{B}'_1}_1)$ and $(\widehat{U}^{\mathcal{F}_1}_1,\widehat{U}^{\mathcal{B}'_1}_1)$ can be used for the $\left(\mathcal{L}_{X|Y}\right)^c$ in the  block communication of time $2$.

Then the following blocks just repeat these operations. Therefore random bits of $\left(\mathcal{L}_{X|Y}\right)^c$ can be conveyed from Alice to Bob separately and securely over the blocks.

\subsection{Strong Security Polar Coding Scheme}

Now we combine this modified multi-block chaining structure with the OTP chain based secure polar coding scheme to construct the strong security polar coding scheme for the non-degraded delay CSI WTC model.

\vbox{}
\textbf{Block $0$:}
\begin{itemize}
\item Legitimate parties obtain the polarized subsets of main channel as $\mathcal{L}_{X|Y}$ and $\left(\mathcal{L}_{X|Y}\right)^c$;
\item Assigning the $u_0^N$ for polar coding:
\begin{itemize}
\item uniformly distributed random bits are assigned to subset $\mathcal{L}_{X|Y}$;
\item pre-shared and secure frozen bits are assigned to subset $\left(\mathcal{L}_{X|Y}\right)^c$;
\end{itemize}
\item Alice encodes $u^N_0$ into the optimally distributed channel input $x^N_0$ by polar encoding $x^N_0=u^N_0\mathbf{G}_N$, and transmits $x^N_0$ to Bob over the known and fixed main channel block;
\item Bob receives $y^N_0$ and decodes it into the estimated $\hat{u}^N_0$ by using the SC decoding:
\begin{equation}
\hat{u}_0^i=
\begin{cases}
  \arg \max \limits_{u\in\left\{0,1\right\}}p_{U^i|U^{1:i-1}Y^N}\left(u|\hat{u}^{1:i-1}_0y^N_0\right)\text{, if } i\in\mathcal{L}_{X|Y} \\
  \text{pre-shared secure frozen bit, if }i\in\left(\mathcal{L}_{X|Y}\right)^c
\end{cases}
\end{equation}
\item After the block communication, with the delay CSI realization $\mathbf{s}_0$.
\begin{itemize}
\item Alice identifies $u^{\mathcal{I}'_0}_0$ as the key stream for next block encryption, also identifies $u^{\mathcal{F}_0}_0$ and $u^{\mathcal{B}'_0}_0$ as the random bits for assigning the $u_1^{\left(\mathcal{L}_{X|Y}\right)^c}$ in the next block;
\item Bob identifies $\hat{u}^{\mathcal{I}'_0}_0$  from the decoded message as the key stream for next block decryption, also identifies $u^{\mathcal{F}_0}_0$ and $\hat{u}^{\mathcal{B}'_0}_0$ as the random bits for decoding the $u_1^{\left(\mathcal{L}_{X|Y}\right)^c}$ in the next block;
\end{itemize}
\end{itemize}

\vbox{}
\textbf{Block $t$, $t\in[\![1,T]\!]$:}
\begin{itemize}
\item Legitimate parties obtain the divided subsets of last block as $(\mathcal{I}'_{t-1},\mathcal{B}'_{t-1},\mathcal{R}_{t-1},\mathcal{B}_{t-1},\mathcal{F}_{t-1})$ by the CSI realization $s_{t-1}$;
\item Assume a message $M_t$ that satisfies $|M_t|=|\mathcal{I}'_{t-1}|$. Then encrypt $|M_t|$ into ciphertext $E_t$ by $E_t=M_t\oplus u^{\mathcal{I}'_{t-1}}_{t-1}$;
\item Assigning the $u_t^N$ for polar coding:
\begin{itemize}
\item ciphertext $E_t$ is assigned to subset $\mathcal{I}'_{t-1}$;
\item uniformly distributed random bits are assigned to subset $\mathcal{R}_{t-1}$;
\item random bits of $u^{\mathcal{F}_{t-1}}_{t-1}$ and $u^{\mathcal{B}'_{t-1}}_{t-1}$ are assigned to subset $\left(\mathcal{L}_{X|Y}\right)^c$;
\end{itemize}
\item Alice encodes $u^N_t$ into the optimally distributed channel input $x^N_t$ by polar encoding $x^N_t=u^N_t\mathbf{G}_N$, and transmit $x^N_t$ to Bob over the main channel;
\item Bob receives $y^N_t$ and decodes it into the estimated $\hat{u}^N_t$ by using the SC decoding:
\begin{equation}
\hat{u}_t^i=
\begin{cases}
  \arg \max \limits_{u\in\left\{0,1\right\}}p_{U^i|U^{1:i-1}Y^N}\left(u|\hat{u}^{1:i-1}_ty^N_t\right)\text{, if } i\in\mathcal{L}_{X|Y} \\
  \text{corresponding bit in }\hat{u}^{\mathcal{F}_{t-1}}_{t-1} \text{~and~} \hat{u}^{\mathcal{B}'_{t-1}}_{t-1}\text{, if }i\in\left(\mathcal{L}_{X|Y}\right)^c
\end{cases}
\end{equation}
\item Bob extracts $\hat{u}^{\mathcal{I}'_{t-1}}_t$ as the ciphertext and decrypts it by $\widehat{M}_t=\hat{u}^{\mathcal{I}'_{t-1}}_t\oplus \hat{u}^{\mathcal{I}'_{t-1}}_{t-1}$;
\item After the block communication, with the delay CSI realization $\mathbf{s}_t$.
\begin{itemize}
\item Alice identifies $u^{\mathcal{I}'_t}_t$ as the key stream for next block encryption, also identifies $u^{\mathcal{F}_t}_t$ and $u^{\mathcal{B}'_t}_t$ as the random bits for assigning the $u_{t+1}^{\left(\mathcal{L}_{X|Y}\right)^c}$ in the next block;
\item Bob identifies $\hat{u}^{\mathcal{I}'_t}_t$  from the decoded message as the key stream for next block decryption, also identifies $\hat{u}^{\mathcal{F}_t}_t$ and $\hat{u}^{\mathcal{B}'_t}_t$ as the random bits for decoding the $\hat{u}_{t+1}^{\left(\mathcal{L}_{X|Y}\right)^c}$ in the next block;
\end{itemize}
\end{itemize}

%\IEEEpubidadjcol

\subsection{Performance Analysis}

Now we analyze the performance of the proposed strong security polar coding scheme and theoretically discuss its reliability, security and secrecy rate under the delay CSI assumption.

\subsubsection{Reliability}  reliability of the proposed strong security polar coding scheme is on the error probability of decoding the ciphertext, key stream and the random bits of subset $\mathcal{B}'$ from time $0$ to time $T$.

\begin{proposition}\label{prop_strong_reliability} The proposed strong security polar coding scheme achieves reliability over the delay CSI WTC model.
\begin{IEEEproof} Similar as in Proposition~\ref{prop_weak_reliability}, for the error probability of entire $T+1$ block communication, have
\begin{equation}
\begin{split}
\mathrm{P_e}(T+1)=&\sum_{t=1}^T \sum_{i\in\mathcal{I}'_{t-1}}Z(U^i|U^{1:i-1},Y^N)+\sum_{t=0}^{T-1} \sum_{i\in\mathcal{I}'_t}Z(U^i|U^{1:i-1},Y^N)\\
&+\sum_{t=0}^{T-1} \sum_{i\in\mathcal{B}'_t}Z(U^i|U^{1:i-1},Y^N)\\
=&3TO(2^{-N^\beta}),
\end{split}
\end{equation}
which proves the reliability.
\end{IEEEproof}
\end{proposition}

\subsubsection{Strong security} In the proposed strong security polar coding scheme, key streams are carried by $U^{\mathcal{I}'_t}_t$ while ciphertexts are carried by $U^{\mathcal{I}'_{t-1}}_t$. Thus for the entire $T+1$ times block communication, the strong security can be measured by the overall information leakage of all the subset $\mathcal{I}'$ from time $0$ to time $T$.

\begin{definition}
For arbitrary subset $\mathcal{A}$ of index $[\![1,N]\!]$, define $\mathrm{a}_1<\mathrm{a}_2<...<\mathrm{a}_{|\mathcal{A}|}$ the corresponding indices of the elements $U^\mathcal{A}$, and
\begin{equation}
U^\mathcal{A}\triangleq U^{\mathrm{a}_1:\mathrm{a}_{|\mathcal{A}|}}=U^{\mathrm{a}_1},U^{\mathrm{a}_1},...,U^{\mathrm{a}_{|\mathcal{A}|}}.
\end{equation}
\end{definition}

%\begin{lemma}(\cite{Gulcu2015})\label{lem_leakage} Considering the polar subset division of index $N$, for arbitrary $\mathcal{A}\subseteq \mathcal{H}_{X|Z}$, $i\in[\![1,|\mathcal{A}|]\!]$, $\mathrm{a}_i\in \mathcal{A}$, $\beta\in\left(0,1/2\right)$, $\delta_{N}=2^{-N^\beta}$, if $\mathcal{A}^c$ is for random bits, then have
%\begin{equation}
%\begin{split}
%H(U_{\mathrm{a}_i}|U^{\mathrm{a}_1:\mathrm{a}_{i-1}},Z^N)&\geq H(U_{\mathrm{a}_i}|U^{1:\mathrm{a}_i-1},Z^N)\\
%&\geq 1-O(N^2 2^{-N^\beta}).
%\end{split}
%\end{equation}
%\end{lemma}

\begin{proposition}\label{prop_strong_security} The proposed strong security polar coding scheme achieves strong security over the non-degraded delay CSI WTC model.
\begin{IEEEproof}
For block $t$, denote $\mathbf{I}^t=U^{\mathcal{I}'_t}_t$, $\mathbf{B}^t=U^{\mathcal{B}'_t}_t$, $\mathbf{F}^t=U^{\mathcal{F}_t}_t$ and $\mathbf{Z}^t=Z_{\mathbf{s}_t}^N$. Then for the entire $T+1$ times block communication, the general information leakage is
\begin{equation}
\mathrm{L}(T+1)=I(\mathbf{I}^{1:T};\mathbf{Z}^{1:T}).
\end{equation}

Now we perform a similar analysis operation as in \cite{Vard2013strong} on the $\mathrm{L}(T+1)$ for the modified multi-block chaining structure. Let
\begin{equation}
\mathfrak{I}_{T}=I(\mathbf{I}^{1:T},\mathbf{B}^T,\mathbf{F}^{T};\mathbf{Z}^{1:T})\geq\mathrm{L}(T+1),
\end{equation}
then for $t\in[\![1,T]\!]$, have
\begin{equation}
\begin{split}
\mathfrak{I}_{t}&=I(\mathbf{I}^{1:t},\mathbf{B}^t,\mathbf{F}^{t};\mathbf{Z}^{1:t})\\
&=I(\mathbf{I}^{1:t},\mathbf{B}^{t},\mathbf{F}^t;\mathbf{Z}^t)+I(\mathbf{I}^{1:t},\mathbf{B}^{t},\mathbf{F}^t;\mathbf{Z}^{1:t-1}|\mathbf{Z}^t)\\
&\overset{(a)}{=}I(\mathbf{I}^t,\mathbf{B}^t,\mathbf{F}^t;\mathbf{Z}^t)+I(\mathbf{I}^{1:t},\mathbf{B}^{t},\mathbf{F}^t;\mathbf{Z}^{1:t-1}|\mathbf{Z}^t)\\
&\leq I(\mathbf{I}^t,\mathbf{B}^t,\mathbf{F}^t;\mathbf{Z}^t)+I(\mathbf{I}^{t},\mathbf{B}^{t-1:t},\mathbf{F}^{t-1:t},\mathbf{Z}^t;\mathbf{Z}^{1:t-1})\\
&\overset{(b)}{=}I(\mathbf{I}^t,\mathbf{B}^t,\mathbf{F}^t;\mathbf{Z}^t)+I(\mathbf{I}^{1:t-1},\mathbf{B}^{t-1},\mathbf{F}^{t-1};\mathbf{Z}^{1:t-1})\\
&=I(\mathbf{I}^t,\mathbf{B}^t,\mathbf{F}^t;\mathbf{Z}^t)+\mathfrak{I}_{t-1},
\end{split}
\end{equation}
where $(a)$ is due to Markov chain
\begin{equation}
\mathbf{I}^{1:t-1}- \mathbf{I}^t,\mathbf{B}^t,\mathbf{F}^t-\mathbf{Z}^t,
\end{equation}
and $(b)$ is due to Markov chain
\begin{equation}
\mathbf{I}^t,\mathbf{B}^t,\mathbf{F}^t,\mathbf{Z}^t- \mathbf{I}^{1:t-1},\mathbf{B}^{t-1},\mathbf{F}^{t-1} - \mathbf{Z}^{1:t-1}.
\end{equation}

Since Eve do not know the initially pre-shared frozen bits for $\left(\mathcal{L}_{X|Y}\right)^c$ at time $0$, have
\begin{equation}
\mathrm{L}(T+1)\leq\mathfrak{I}_T\leq\sum_{t=0}^T I(\mathbf{I}^t,\mathbf{B}^t,\mathbf{F}^t;\mathbf{Z}^t).
\end{equation}
Also because $\mathcal{R}_t$ are set for transmitting random bits, have
\begin{equation}
\begin{split}
I(\mathbf{I}^t,\mathbf{B}^t,\mathbf{F}^t;\mathbf{Z}^t)=&I(U^{\mathcal{I}'_t\cup\mathcal{B}'_t\cup\mathcal{F}_t};Z^N_{\mathbf{s}_t})=\sum_{i=1}^{|\mathcal{I}'_t\cup\mathcal{B}'_t\cup\mathcal{F}_t|}I(U^{\mathrm{a}_i};Z^N_{\mathbf{s}_t}|U^{\mathrm{a}_1:\mathrm{a}_{i-1}})\\
\overset{(a)}{=}& \sum_{i=1}^{|\mathcal{I}'_t\cup\mathcal{B}'_t\cup\mathcal{F}_t|}I(U^{\mathrm{a}_i};U^{\mathrm{a}_1:\mathrm{a}_{i-1}},Z^N_{\mathbf{s}_t})\\
\leq&\sum_{i=1}^{|\mathcal{I}'_t\cup\mathcal{B}'_t\cup\mathcal{F}_t|}I(U^{\mathrm{a}_i};U^{1:\mathrm{a}_{i}-1},Z^N_{\mathbf{s}_t})\\
=&\sum_{i=1}^{|\mathcal{I}'_t\cup\mathcal{B}'_t\cup\mathcal{F}_t|}\left[H(U^{\mathrm{a}_i})-H(U^{\mathrm{a}_i}|Z^N_{\mathbf{s}_t},U^{1:\mathrm{a}_{i}-1})\right]\overset{(b)}{\leq}O(N 2^{-N^\beta}),
\end{split}
\end{equation}
where $(a)$ is because each $U^{\mathrm{a}_{i}}$ is independent; $(b)$ is due to $\mathcal{I}'_t\cup\mathcal{B}'_t\cup\mathcal{F}_t= \mathcal{H}_{X|Z}^{(\mathbf{s}_t)}$, $Z(X|Y)^2\leq H(X|Y)$ and $\mathcal{H}_{X|Z}^{(\mathbf{s}_t)}=\left\{j\in[\![1,N]\!]:Z\left(U^j|U^{1:j-1},Z^N_{\mathbf{s}_t}\right)\geq1-\delta_N \right\}$. Therefore, we finally have
\begin{equation}
\mathrm{L}(T+1)\leq (T+1)O(N 2^{-N^\beta}),
\end{equation}
which proves the strong security.
\end{IEEEproof}
\end{proposition}

\subsubsection{Secrecy rate} Now we discuss the achievable secrecy rate under the reliability and strong security criterions.

\begin{proposition}\label{prop_strong_rate} Over the non-degraded delay CSI WTC model, with a large enough $T$, the achievable secrecy rate of the proposed strong security polar coding scheme can approach the secrecy capacity of perfect CSI case.
\begin{IEEEproof}
According to the proposed strong security polar coding scheme, ciphertext are carried by $U_t^{\mathcal{I}'_{t-1}}$ for $t\in[\![1,T]\!]$, hence for the secrecy rate, have
\begin{equation}
\begin{split}
\mathrm{R_s}(T+1)&=\frac{1}{N(T+1)}\sum_{t=1}^T|\mathcal{I}'_{t-1}|=\frac{1}{N(T+1)}\sum_{t=1}^T(|\mathcal{I}_{t-1}|-|\mathcal{B}'_{t-1}|)\\
&=\frac{1}{T+1}\sum_{t=1}^T\frac{|\mathcal{I}_{t-1}\cup\mathcal{R}_{t-1}|-|\mathcal{B}_{t-1}\cup\mathcal{R}_{t-1}|}{N}.
\end{split}
\end{equation}
According to \eqref{eq_rmk1}, have
\begin{equation}
\lim_{N\rightarrow\infty}\frac{|\mathcal{I}_{t-1}\cup\mathcal{R}_{t-1}|-|\mathcal{B}_{t-1}\cup\mathcal{R}_{t-1}|}{N}=I(U;Y)-\lim_{N\rightarrow\infty}\frac{1}{N}I(U^N;Z^N_{\mathbf{s}_{t-1}}).
\end{equation}
Thus have
\begin{equation}
\lim_{N\rightarrow\infty}\mathrm{R_s}(T+1)=\frac{1}{T+1}\sum_{t=1}^T\left[ I(U;Y)-\lim_{N\rightarrow\infty}\frac{1}{N}I(U^N;Z^N_{\mathbf{s}_{t-1}}) \right],
\end{equation}
Then by comparing with \eqref{eq_perfect_cb} and \eqref{eq_perfect_ca}, we can observe that $I(U;Y)-\lim_{N\rightarrow\infty}\frac{1}{N}I(U^N;Z_\mathbf{s_{t-1}}^N)$ can reach the perfect CSI average secrecy capacity of a block with state $\mathbf{s_{t-1}}$. Thus with a large enough $T$, we have
\begin{equation}
\lim_{N\rightarrow\infty}\mathrm{R_s}(T+1)\approx I(U;Y)-\lim_{N\rightarrow\infty}\frac{1}{N}I(U^N;Z^N_{\mathbf{s}}).
\end{equation}
which indicates that under the delay CSI assumption, the achievable secrecy rate of the proposed strong security polar coding scheme can approach the average secrecy capacity of the perfect CSI case with a sufficiently large $T$.
\end{IEEEproof}
\end{proposition}

\section{Simulation of the Schemes}\label{sec_simu}

In this section we present simulations of the proposed secure polar coding schemes.

\subsection{Simulation of the Weak Security Scheme}

First, we setup a precise degraded delay CSI WTC model. Specifically, the main channel is  a binary erase channel (BEC) with erase probability $\epsilon_m=0.1$; the degraded wiretap channel is varying BEC (block varying and arbitrarily varying) with uncertainty erase set $\{\epsilon_{w1}=0.4,\epsilon_{w2}=0.5\}$.

\begin{figure*}[!t]
\centering
\subfloat[Upper bound of legitimate BER]{\includegraphics[width=8.2cm]{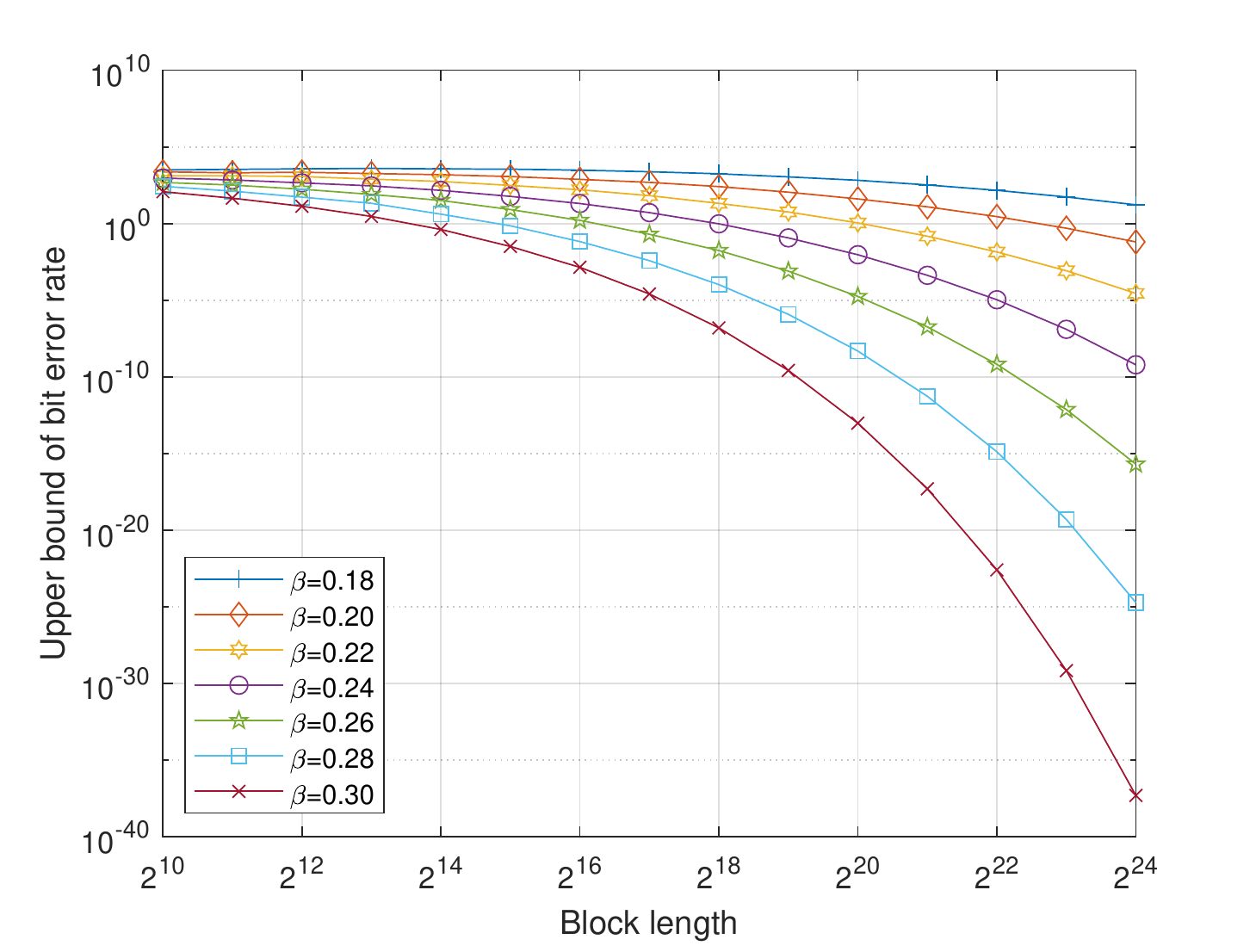}%
\label{fig_weak_upber}}
\hfil
\subfloat[Information leakage]{\includegraphics[width=8.2cm]{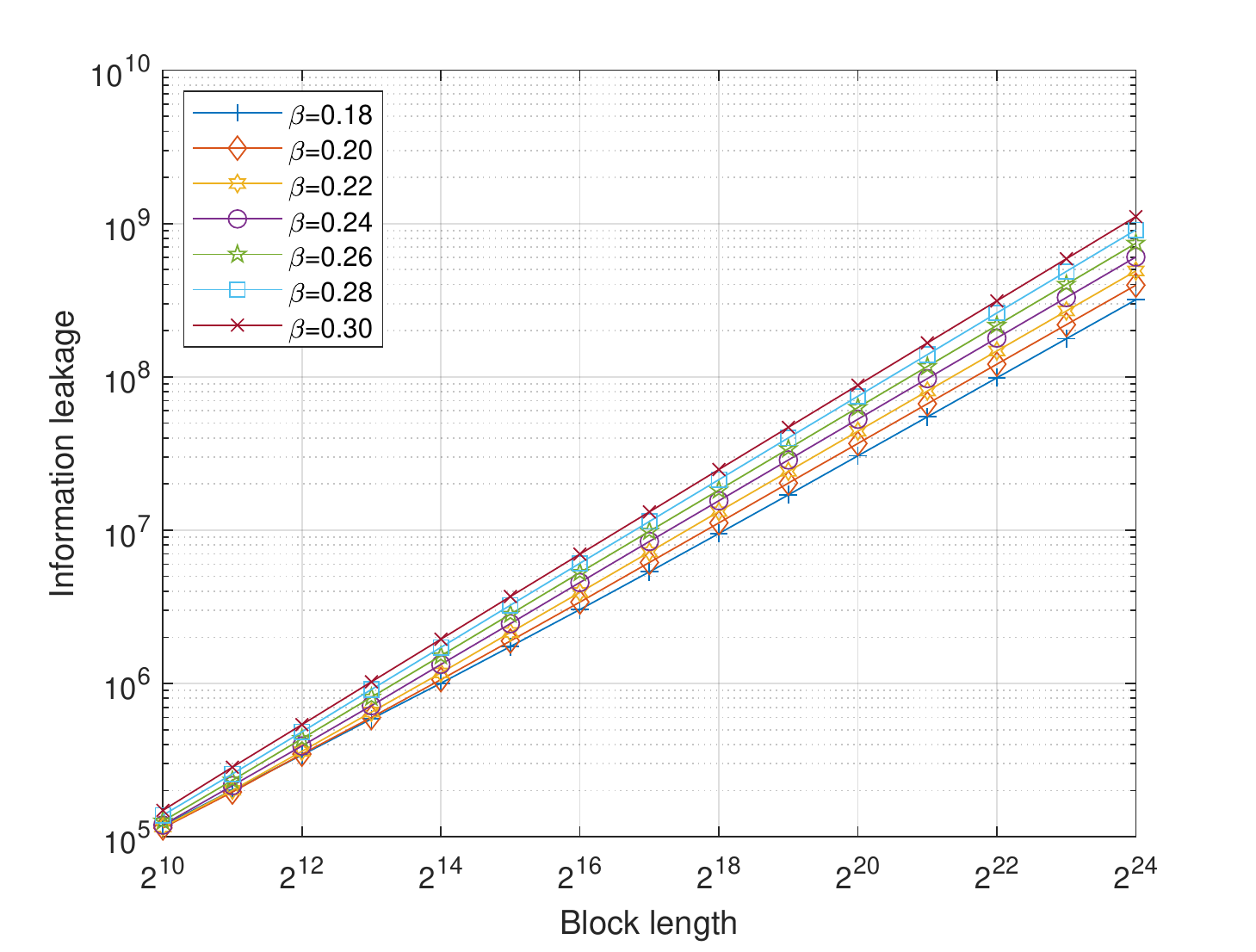}%
\label{fig_weak_leakage}}
\hfil
\subfloat[Information leakage rate]{\includegraphics[width=8.2cm]{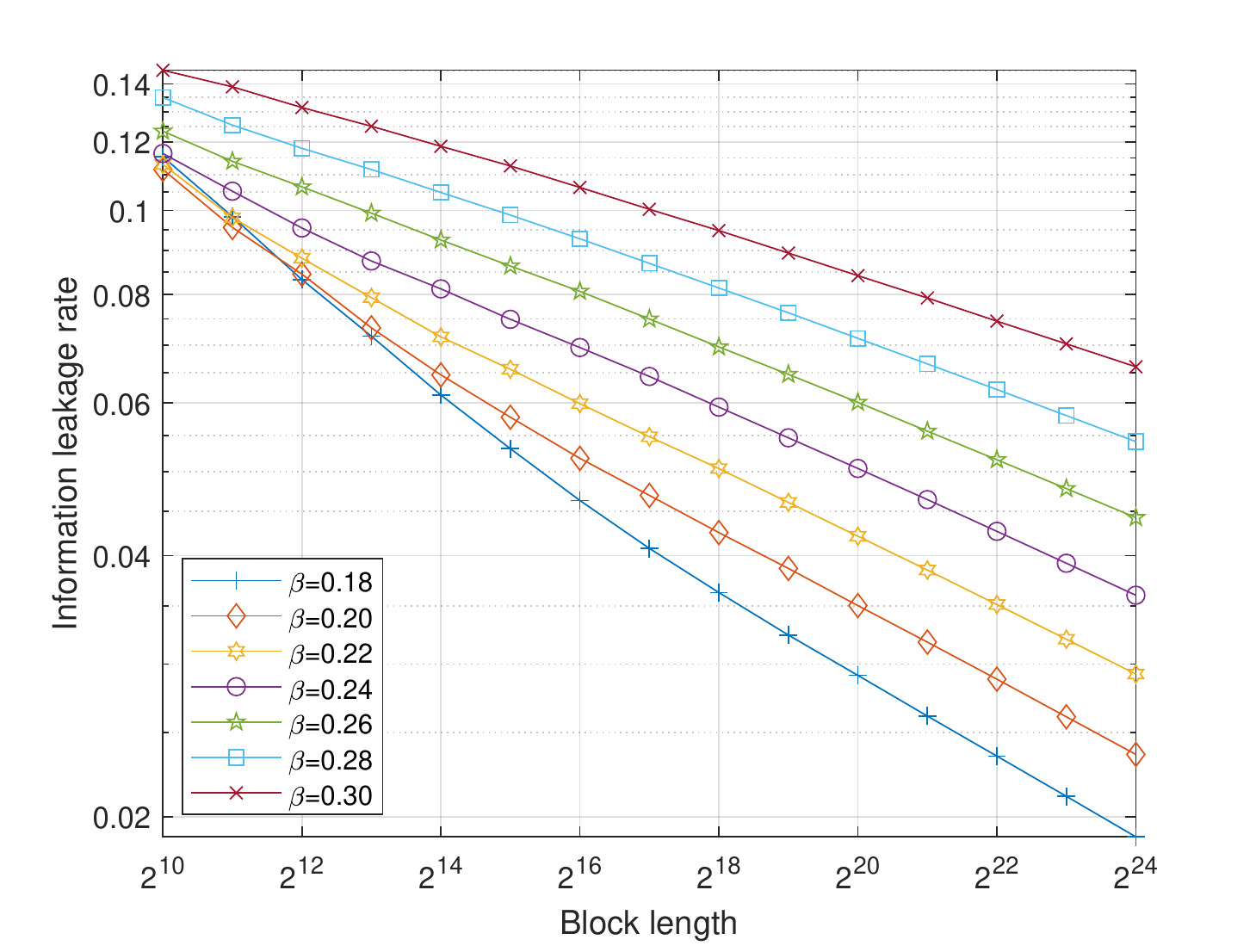}%
\label{fig_weak_leakage_rate}}
\hfil
\subfloat[Secrecy rate]{\includegraphics[width=8.2cm]{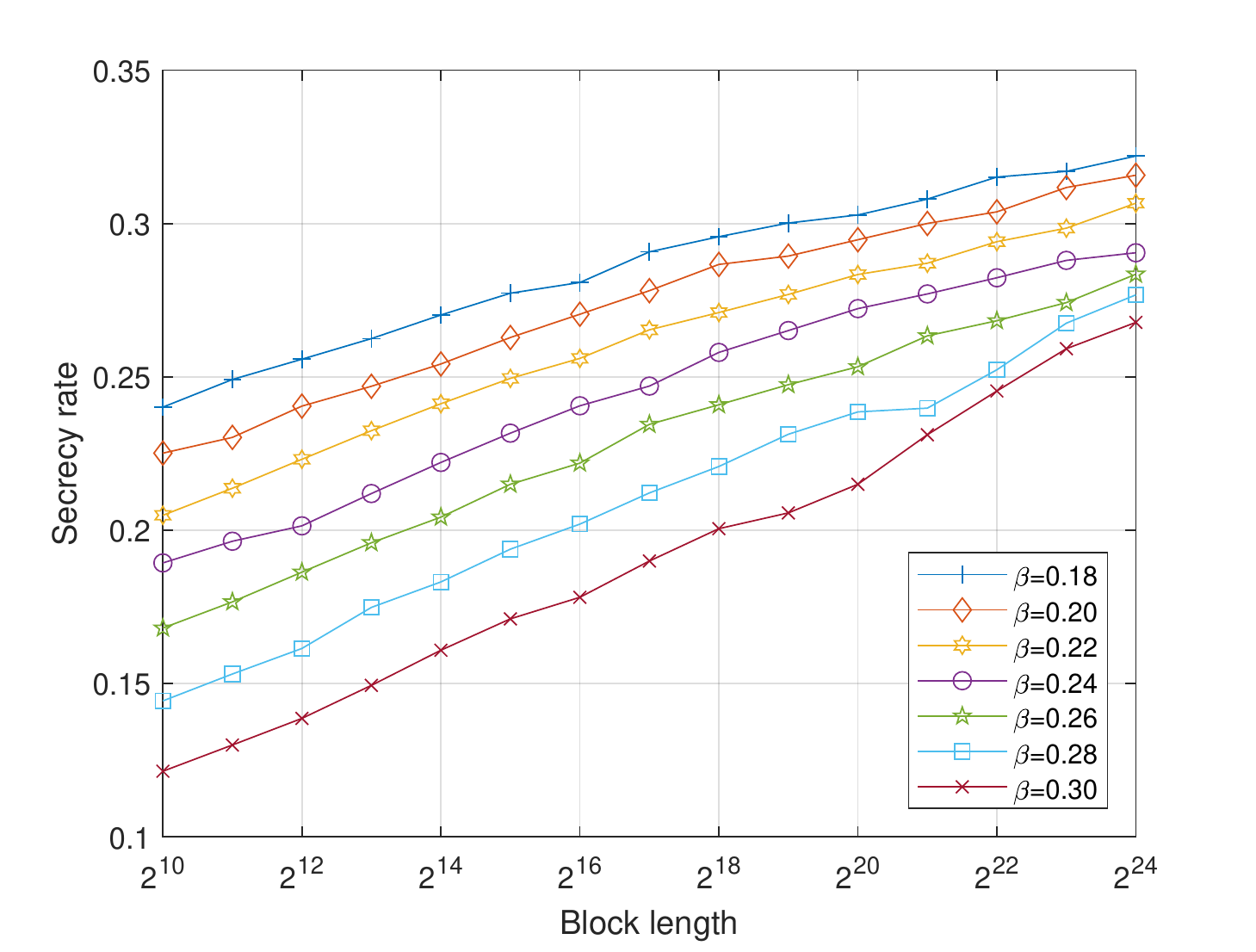}%
\label{fig_weak_rate}}
\caption{Simulation results for the OTP based secure polar coding scheme with $T=1000$.}
\label{fig_weak_theo}
\end{figure*}

Then on this model we carry out the performance stimulation of the weak security coding scheme by choosing $T=1000$, $\beta$ from $0.18$ to $0.30$ and $N$ from $2^{10}$ to $2^{24}$. Results of the performance stimulation are illustrated Fig.~\ref{fig_weak_theo}. Note that all these results are the sum of entire $T+1$ blocks. Fig.~\ref{fig_weak_upber} shows that  when the block length $N$ increases, the theoretical upper bounds of the legitimate BER is vanishing, which meets the Proportion~\ref{prop_weak_reliability} for reliability. Fig.~\ref{fig_weak_leakage} and Fig.~\ref{fig_weak_leakage_rate} shows that the information leakage is increasing but the information leakage rate is vanishing when $N$ increases, which meets the Proportion~\ref{prop_weak_security} for weak security. Fig.~\ref{fig_weak_rate} shows the tendency of secrecy rate approaching the perfect CSI secrecy capacity $0.35$ with an increasing $N$, which meets the Proportion~\ref{prop_weak_rate} for secrecy rate. Also, by comparing the performance of different $\beta$, we can observe that when the parameter $\beta$ gets larger, the weak security scheme can obtain a better legitimate BER, but a worse information leakage rate and secrecy rate, which indicates the significance of parameter $\delta_N=2^{-N^{\beta}}$ for finite block length secure polar codes.

\begin{figure*}[!t]
\centering
\subfloat[Experimental legitimate BER]{\includegraphics[width=8.2cm]{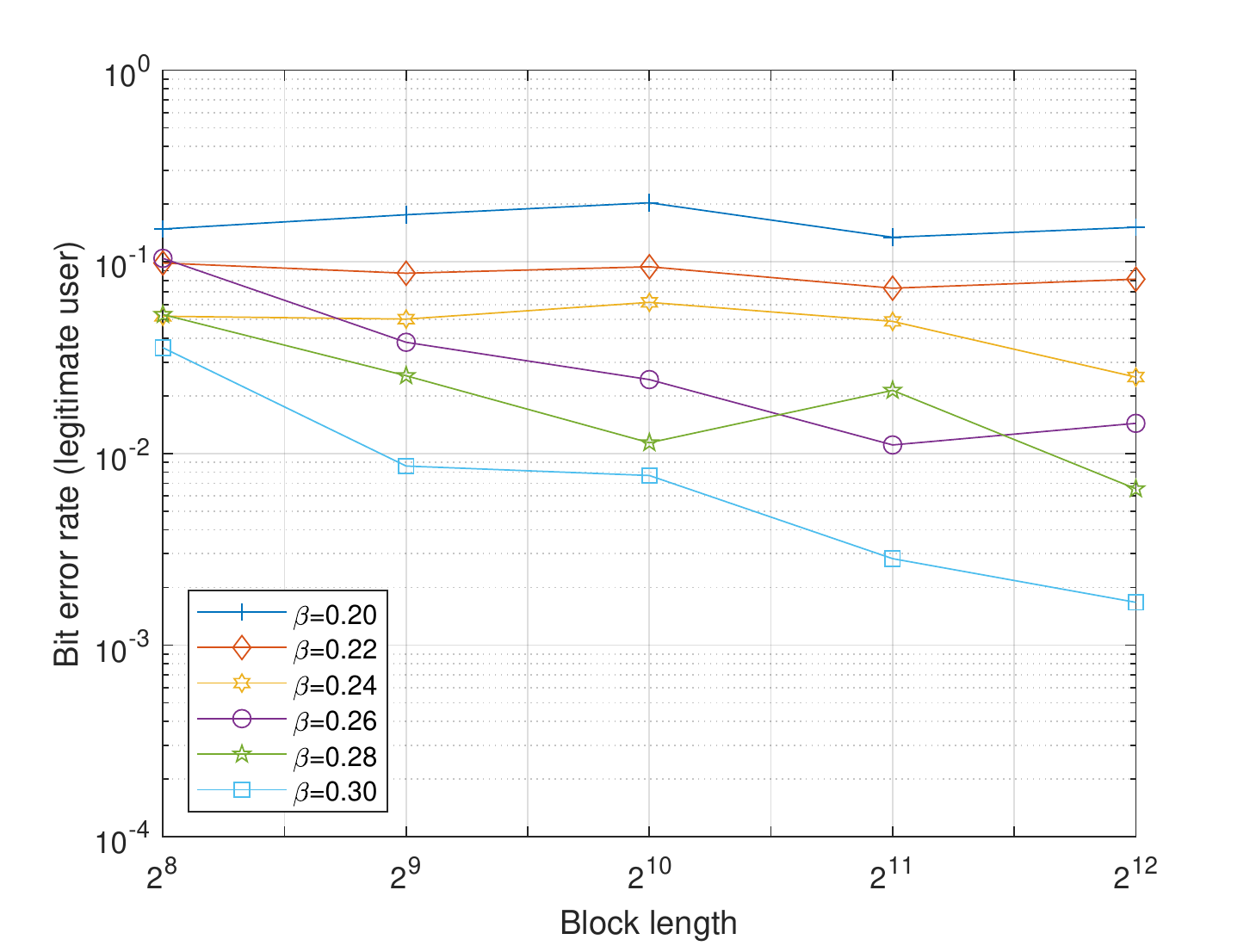}%
\label{fig_weak_lber}}
\hfil
\subfloat[Experimental eavesdropping BER]{\includegraphics[width=8.2cm]{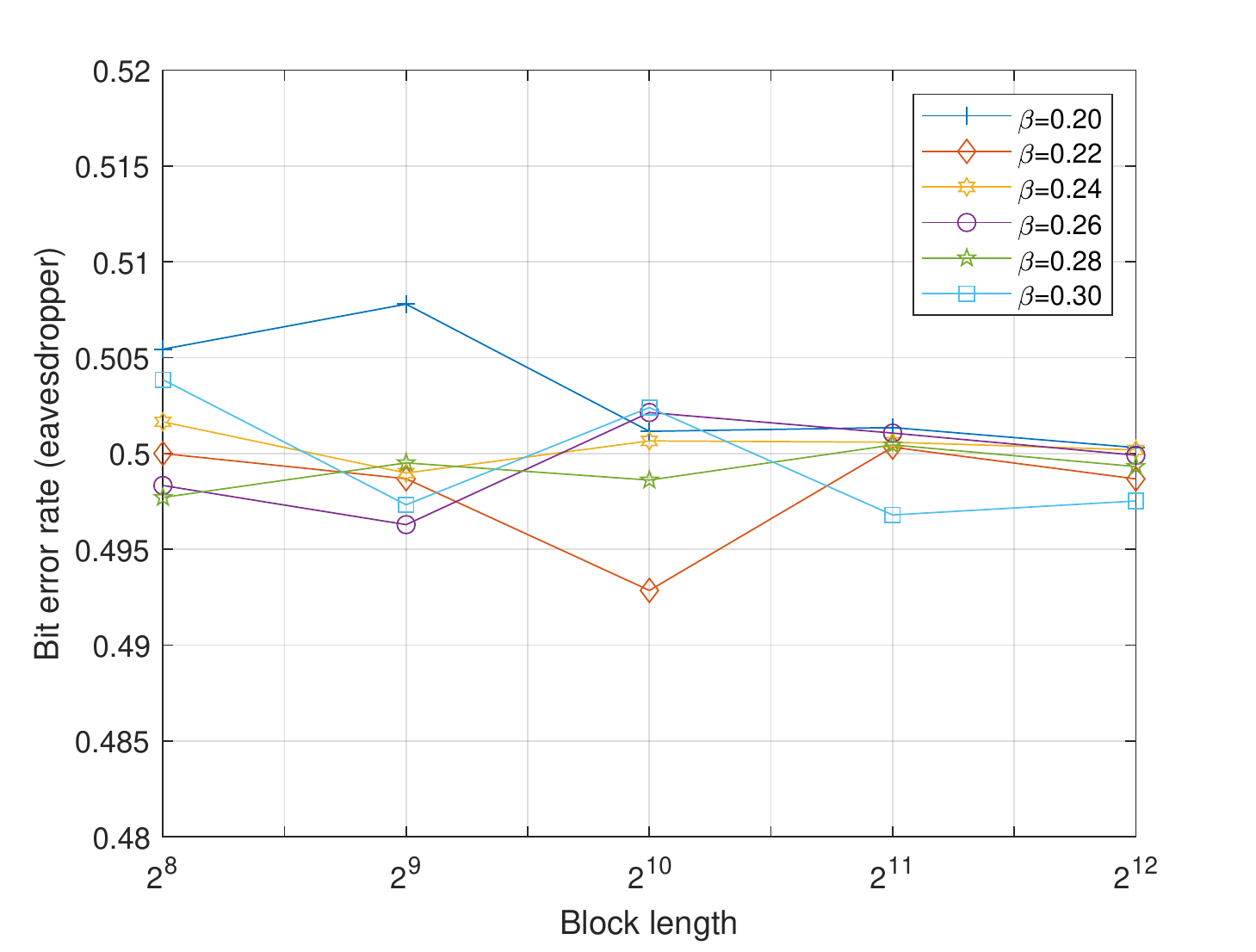}%
\label{fig_weak_eber}}
\caption{Experimental BER results of weak security communication test with $T=3$.}
\label{fig_weak_sim}
\end{figure*}

We also setup a test for the weak security communication process with $T=3$, $\beta$ from $0.20$ to $0.30$ and $N$ from $2^{8}$ to $2^{12}$. As illustrated in Fig.~\ref{fig_weak_sim}, when the block length increases, the experimental legitimate BER is dropping to an acceptable low level, and the experimental eavesdropper BER is approaching $0.5$ for uniformly distributed binary confidential message. Thus both reliability and security is obtained.

%\begin{figure*}[!h]
%\centering
%\includegraphics[width=10cm]{zhao4.pdf}
%\caption{Upper bound of BER for legitimate receiver with $T=1000$, including both ciphertext and key stream.}
%\label{fig_weak_upber}
%\end{figure*}
%
%\begin{figure*}[!h]
%\centering
%\includegraphics[width=10cm]{zhao5.pdf}
%\caption{Information leakage of the weak security scheme with $T=1000$.}
%\label{fig_weak_leakage}
%\end{figure*}
%
%\begin{figure*}[!h]
%\centering
%\includegraphics[width=10cm]{zhao6.pdf}
%\caption{Information leakage rate of the weak security scheme with $T=1000$.}
%\label{fig_weak_leakage_rate}
%\end{figure*}
%
%\begin{figure*}[!h]
%\centering
%\includegraphics[width=10cm]{zhao7.pdf}
%\caption{Secrecy rate of the weak security scheme with $T=1000$.}
%\label{fig_weak_rate}
%\end{figure*}

%\begin{figure*}[!h]
%\centering
%\includegraphics[width=10cm]{zhao8.pdf}
%\caption{Experimental BER of confidential message for legitimate receiver with $T=3$.}
%\label{fig_theo_rate}
%\end{figure*}
%
%\begin{figure*}[!h]
%\centering
%\includegraphics[width=10cm]{zhao9.pdf}
%\caption{Experimental BER of confidential message for eavesdropper with $T=3$..}
%\label{fig_theo_rate}
%\end{figure*}

\subsection{Simulation of the Strong Security Scheme}

Note that in degraded WTC model, subset $\mathcal{B}$ unstably exists according to different parameter $\beta$ and block length $N$. Thus to test the effectiveness of strong security scheme on subset $\mathcal{B}$, we built a stable subset $\mathcal{\widetilde{B}}$ on the degraded delay CSI WTC model.

Recall that for the precise degraded model in the previous subsection, we have a BEC main channel with erase probability $\epsilon_m=0.1$ and a varying BEC wiretap channel with uncertainty erase set $\{\epsilon_{w1}=0.4,\epsilon_{w2}=0.5\}$. Then we fix three kinds of $N$ length wiretap channel blocks:
\begin{itemize}
  \item block1: stationary BEC channel block with $\epsilon_{w1}$;
  \item block2: stationary BEC channel block with $\epsilon_{w1}$;
  \item block3: non-stationary BEC channel block with erase probabilities uniformly choosing from $\{\epsilon_{w1},\epsilon_{w2}\}$.
\end{itemize}
And we assume that eavesdropper uniformly chooses wiretap channel blocks from them.

According to \eqref{eq_division}, for each block fixed above, we have subsets $\{\mathcal{I}^{(i)},\mathcal{R}^{(i)},\mathcal{B}^{(i)},\mathcal{F}^{(i)}\}$. Next, we separate a same subset $\mathcal{B}_{add}$ with fixed rate $\mathrm{R}_{add}=0.05$ from each $\mathcal{R}^{(i)}$ and satisfies $\mathcal{B}_{add}\subseteq\mathcal{R}^{(1)}\cap\mathcal{R}^{(2)}\cap\mathcal{R}^{(3)}$. Note that $\mathcal{B}_{add}\subseteq\mathcal{R}^{(i)}\subseteq\left(\mathcal{H}_{X|Z}^{(i)}\right)^c$, so it has a same reliability as the subset $\mathcal{B}^{(i)}$ from eavesdropper's perspective. Also, since bits in subset $\mathcal{B}^{(i)}$ are directly decoded by the decoded bits of previous block, it can be pretended unreliable to legitimate parties (even though it is reliable). Therefore, we can have a stable subset $\mathcal{\widetilde{B}}^{(i)}=\mathcal{B}^{(i)}\cup\mathcal{B}_{add}$ for each block, which plays a same role as the original $\mathcal{B}^{(i)}$. Then we have the adjusted subsets for strong security scheme as $\{\mathcal{\widetilde{L}}_{X|Y}, \mathcal{\widetilde{I}}'^{(i)}, \mathcal{\widetilde{B}}'^{(i)}, \mathcal{\widetilde{R}}^{(i)}, \mathcal{\widetilde{B}}^{(i)}, \mathcal{F}^{(i)}\}$, where $\mathcal{\widetilde{L}}_{X|Y}=\mathcal{L}_{X|Y}\setminus\mathcal{B}_{add}$, $\mathcal{\widetilde{I}}'^{(i)}=\mathcal{I}^{(i)}\setminus\mathcal{\widetilde{B}}'^{(i)}$, $|\mathcal{\widetilde{B}}'^{(i)}|=|\mathcal{\widetilde{B}}^{(i)}|$ and $\mathcal{\widetilde{R}}^{(i)}=\mathcal{R}^{(i)}\setminus\mathcal{B}_{add}$.

\begin{figure*}[!t]
\centering
\subfloat[Upper bound of legitimate BER]{\includegraphics[width=8.2cm]{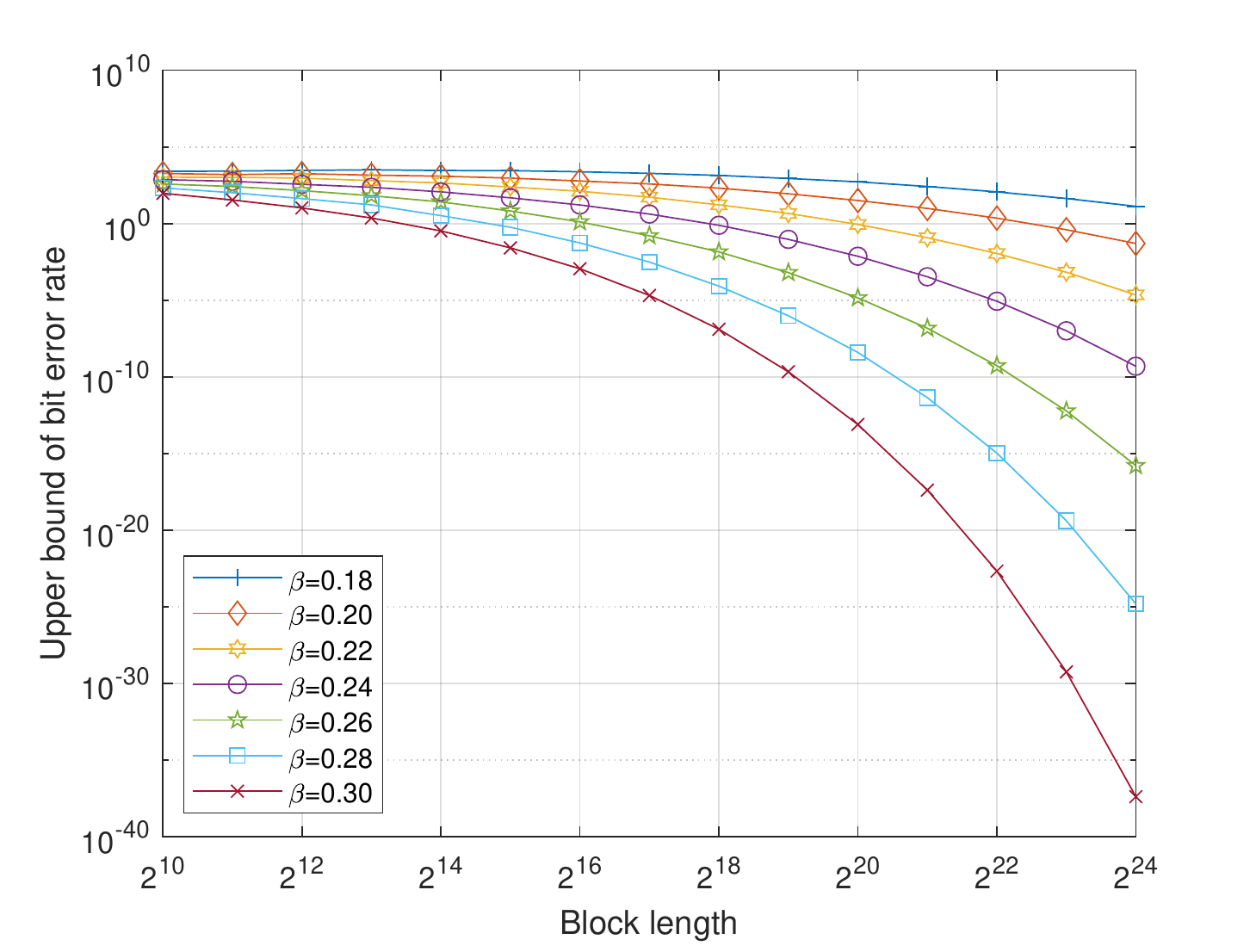}%
\label{fig_strong_upber}}
\hfil
\subfloat[Information leakage]{\includegraphics[width=8.2cm]{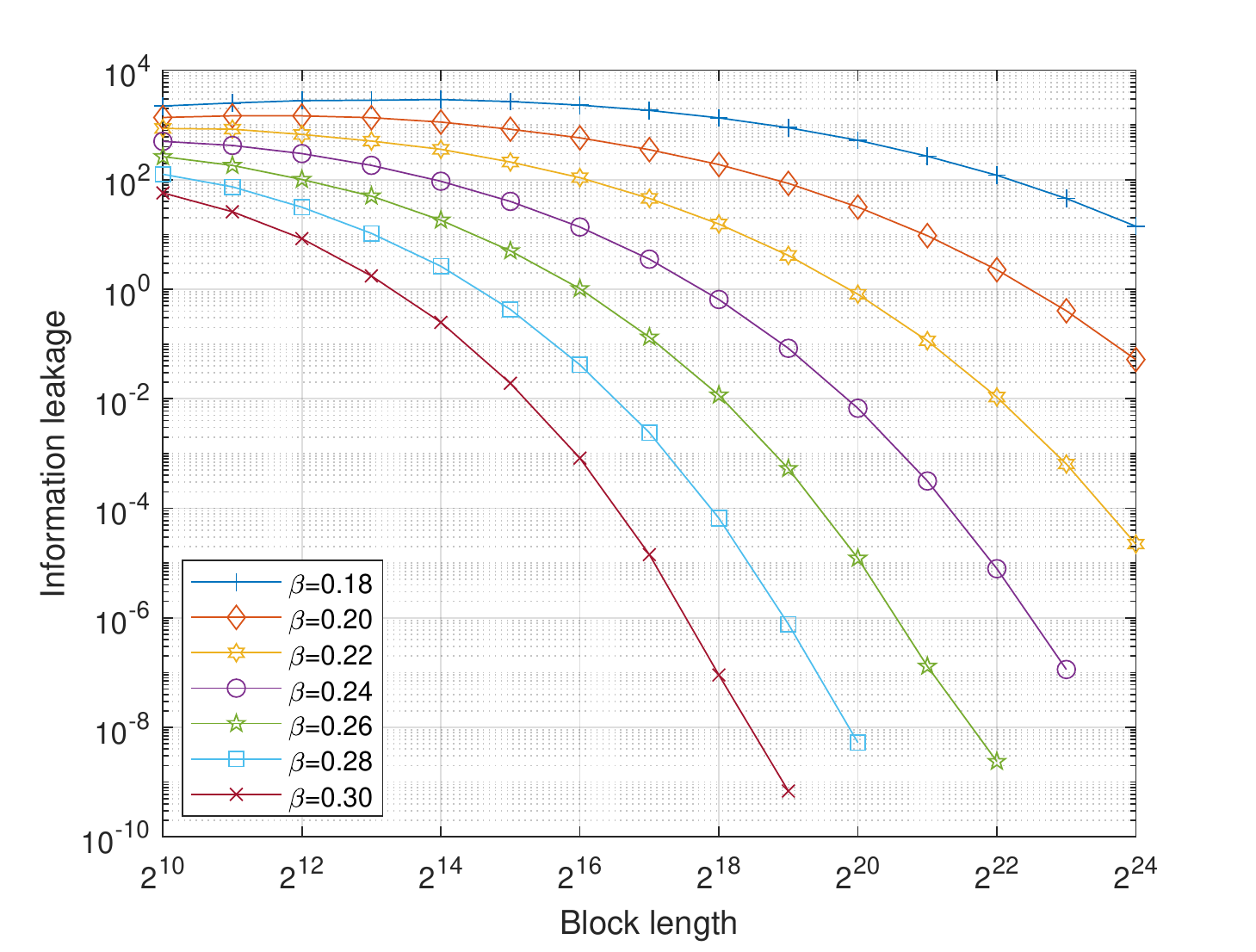}%
\label{fig_strong_leakage}}
\hfil
\subfloat[Secrecy rate]{\includegraphics[width=8.2cm]{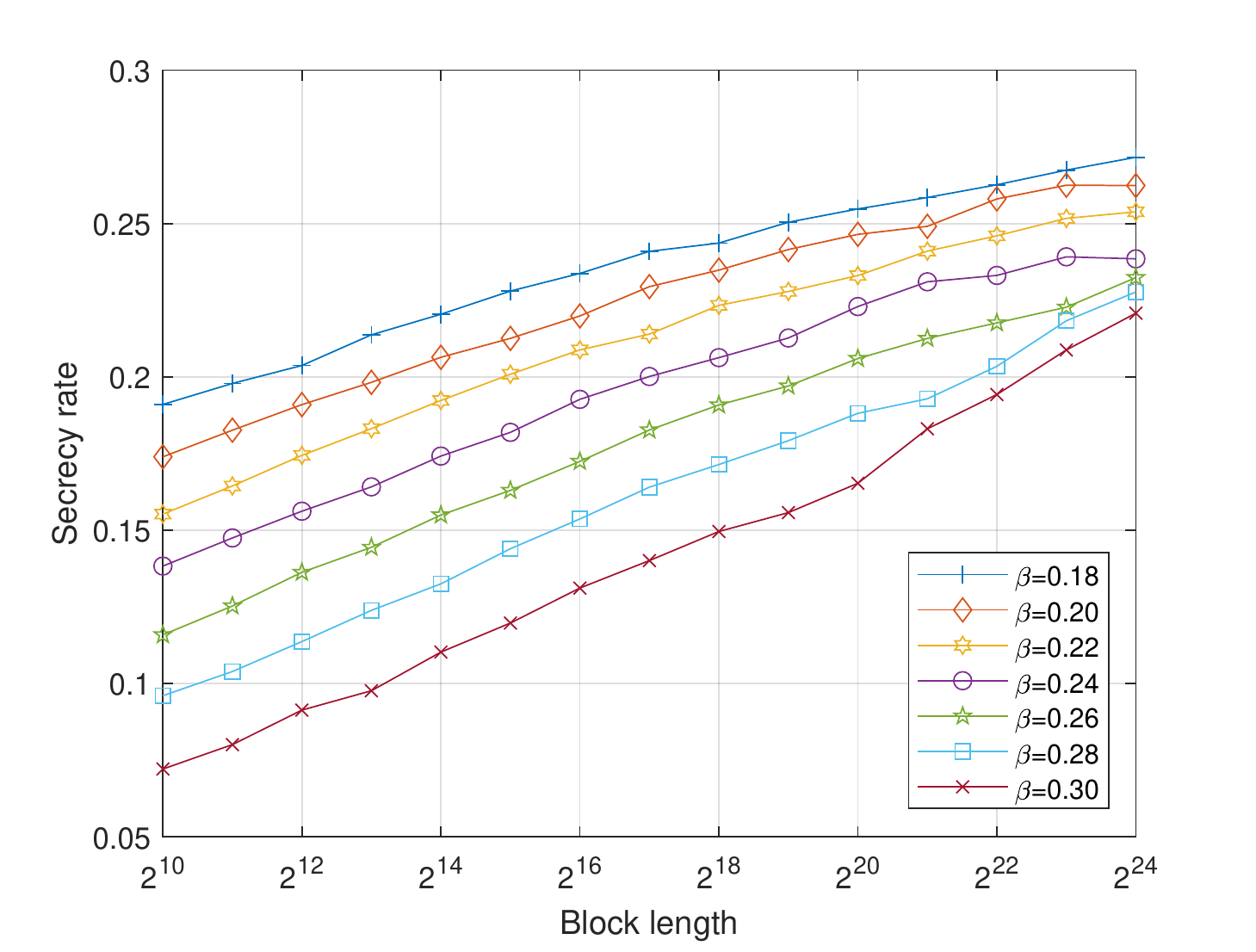}%
\label{fig_strong_rate}}
\caption{Simulation results for the strong security polar coding scheme with $T=1000$.}
\label{fig_strong_theo}
\end{figure*}

Next, based on this constructed model and subsets, we carry out the performance stimulation of the strong security coding scheme by choosing $T=1000$, $\beta$ from $0.18$ to $0.30$ and $N$ from $2^{10}$ to $2^{24}$. Results of the stimulation are illustrated in Fig.~\ref{fig_strong_theo}. Note that all these results of strong security scheme are the sum of entire $T+1$ blocks. Fig.~\ref{fig_strong_upber} shows that when the block length $N$ increases, the theoretical upper bounds of the legitimate BER is vanishing, which meets the Proportion~\ref{prop_strong_reliability} for reliability. Fig.~\ref{fig_strong_leakage} shows that the information leakage rate is vanishing when $N$ increases, which meets the Proportion~\ref{prop_strong_security} for strong security. Fig.~\ref{fig_strong_rate} shows the tendency of secrecy rate approaching the perfect CSI secrecy capacity $0.30$ with an increasing $N$, which meets the Proportion~\ref{prop_strong_rate} for secrecy rate. Also, for the performance of different $\beta$, we can observe that when the parameter $\beta$ gets larger, the strong security scheme can obtain a better legitimate BER and information leakage but worse secrecy rate.

\begin{figure*}[!t]
\centering
\subfloat[Experimental legitimate BER]{\includegraphics[width=8.2cm]{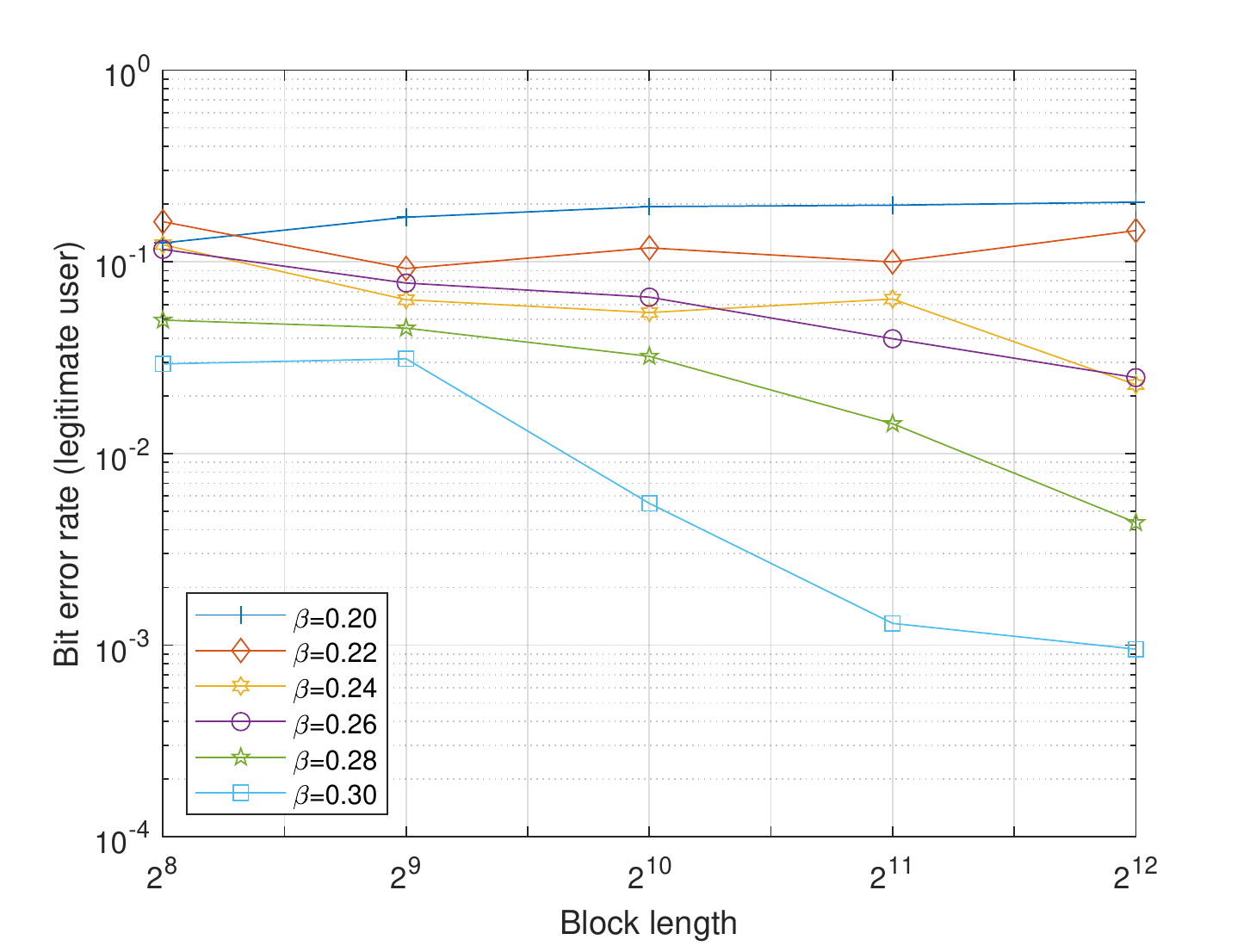}%
\label{fig_strong_lber}}
\hfil
\subfloat[Experimental eavesdropping BER]{\includegraphics[width=8.2cm]{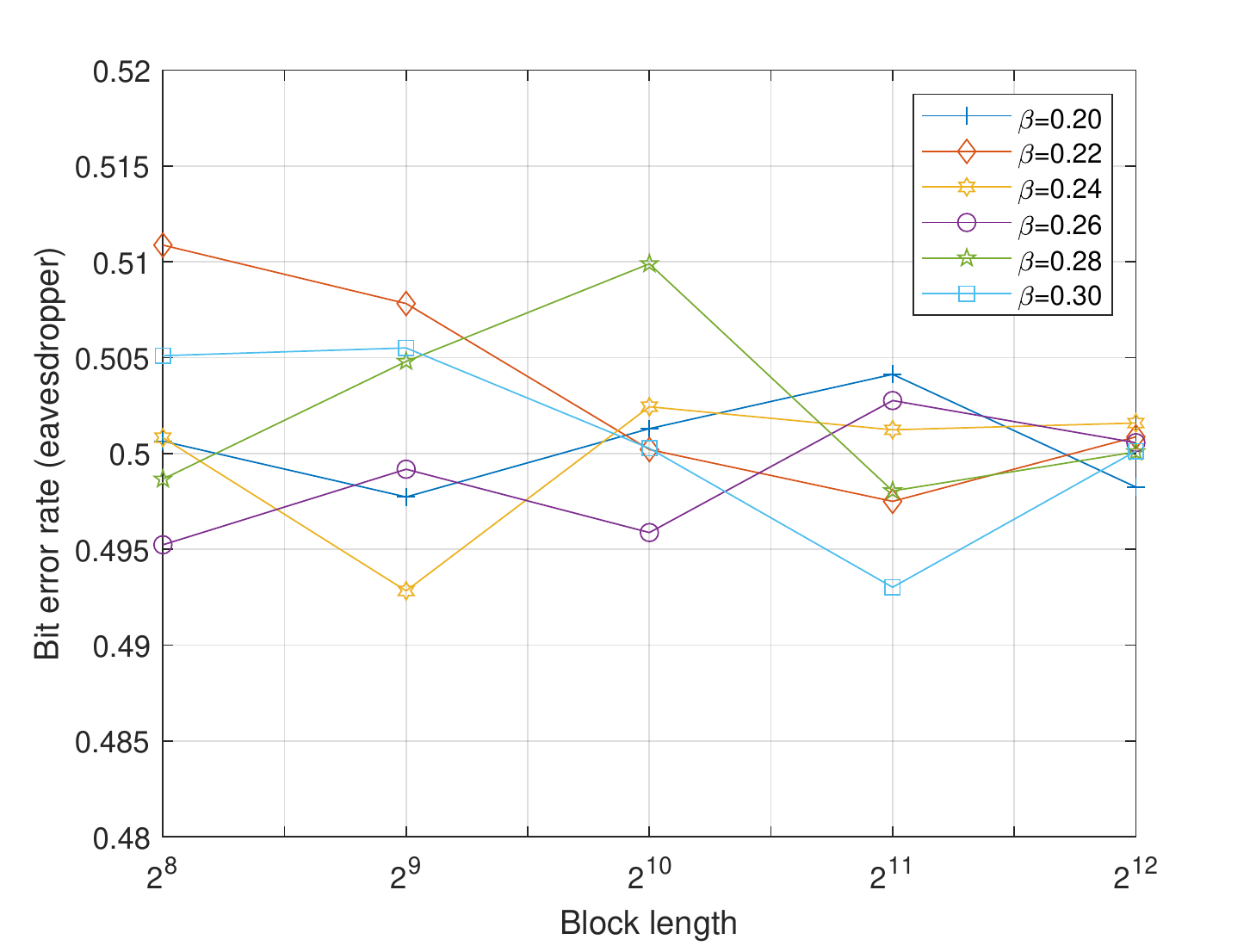}%
\label{fig_strong_eber}}
\caption{Experimental BER results of strong security communication test with $T=3$.}
\label{fig_strong_sim}
\end{figure*}

Further, we perform a communication test for the strong security scheme with $T=3$, $\beta$ from $0.20$ to $0.30$ and $N$ from $2^{8}$ to $2^{12}$. As illustrated in Fig.~\ref{fig_strong_sim}, when the block length increases, the experimental legitimate BER is dropping to an acceptable low level, and the experimental eavesdropper BER is approaching $0.5$ for uniformly distributed binary confidential message. Thus both reliability and security is obtained.

\section{Conclusion}\label{sec_con}

In this paper, we have presented a practical delay CSI WTC model with varying wiretap channel, and then investigate the corresponding secure polar coding scheme to achieve a secure and reliable communication.

As the first step, we proposed a OTP chain based secure polar coding scheme with a preliminary solution for the unidentifiable problem of the neither secure nor reliable subset $\mathcal{B}$ that we assign the subset $\mathcal{H}_X\cap \left(\mathcal{L}_{X|Y}\right)^c$ with publicly known frozen bits. However, this preliminary solution can only achieve weak security over degraded delay CSI WTC.

In the aim of achieving both strong security and reliability over non-degraded delay CSI WTC, we have further discussed the remaining limitation for applying the multi-block chaining structure and presented an new solution named as modified multi-block chaining structure which uses the secure subset $\mathcal{F}$ and subset $\mathcal{B}'$ for conveying the bits for $\left(\mathcal{L}_{X|Y}\right)^c$. Finally by combining this modified multi-block chaining structure with weak security coding scheme, we have proposed a strong security polar coding scheme which can approach the secrecy capacity of perfect CSI case only with delay CSI assumption, with both reliability and strong security.

At last, we have carried out stimulations which has verified the performance of both weak and strong security coding schemes.

% or


%\appendices

%\section{}

\section*{Acknowledgment}

This work is supported in part by the Natural Science Foundation of Hubei Province (Grant No.2019CFB137) and the Fundamental Research Funds for the Central Universities
(Grant No.2662017QD042, No.2662018JC007).

\ifCLASSOPTIONcaptionsoff
  \newpage
\fi



%\bibliographystyle{IEEEtran}


\begin{IEEEbiographynophoto}{Yizhi Zhao}
received the Ph.D. degree in the school of Optical and Electronic Information from the Huazhong University of Science and Technology, Wuhan,
China, in 2017.

He is currently an Assistant Professor with the College of Information, Huazhong Agricultural University. His research interests include physical layer security, communication security, VLSI design and machine learning.
\end{IEEEbiographynophoto}

\begin{IEEEbiographynophoto}{Hongmei Chi}
received her Ph.D. degree in the School of Mathematics and Statistics from Wuhan University, Wuhan, China, in 2014.

Currently, she is an Assistant Professor with College of Science, Huazhong Agricultural University. Her research interest is statistic learning, stochastic analysis and information theory.
\end{IEEEbiographynophoto}


%\newpage


%\vfill


\end{document}